\documentstyle[epic,eepic]{article}

\title{Vertex operators in solvable lattice models}

\author{Omar Foda\thanks{Department of Mathematics, University of Melbourne,
                         Parkville, Victoria 3052, Australia.},
        Michio Jimbo\thanks{Department of Mathematics, Faculty of Science,
                            Kyoto University, Kyoto 606, Japan.},
        Tetsuji Miwa\thanks{Research Institute for Mathematical Sciences,
                            Kyoto University, Kyoto 606, Japan.},\cr
        Kei Miki\thanks{Department of Mathematical Science,
                        Osaka University, Toyonaka, Osaka 560, Japan.},
 and Atsushi Nakayashiki\thanks{The Graduate School of Science and Technology,
                                       Kobe University, Rokkodai, Kobe 657,
Japan.}
}

\begin{document}

\maketitle

\begin{abstract}

We formulate the basic properties of q-vertex
operators in the context of the Andrews-Baxter-Forrester
(ABF) series, as an example of face-interaction models, derive
the $q$-difference equations satisfied by their correlation
functions, and establish their connection with representation
theory.  We also discuss the $q$-difference equations of the
Kashiwara-Miwa (KM) series, as an example of edge-interaction
models.

Next, the Ising model--the simplest special case of both
ABF and KM series--is studied in more
detail using the Jordan-Wigner fermions. In particular, all
matrix elements of vertex operators are calculated.

\end{abstract}

\def\goth#1{{#1}} 

\setcounter{section}{0}
\setcounter{secnumdepth}{1}

\newenvironment{eq}{\begin{eqnarray}}{\end{eqnarray}}
\newtheorem{lemma}{Lemma}

\def\pn#1{\phi^{{\rm NS}}_{#1}}
\def\pr#1{\phi^{{\rm R}}_{#1}}
\def\psz{\psi^{{\rm NS}}_0}
\def\pt{\widetilde{\phi}}
\def\pnh{\widehat{\phi}^{\rm NS}}
\def\prh{\widehat{\phi}^{\rm R}}
\def\C{{\bf C}}
\def\Z{{\bf Z}}
\def\Zt{\Z / 2\Z}
\def\sZ{\small {\bf Z}}
\def\Zh{{\bf Z}+{1\over 2}}
\def\sZh{\sZ+{1\over 2}}
\def\H{{\cal H}}
\def\Hn{{\cal H}^{\rm NS}}
\def\Hr{{\cal H}^{\rm R}}
\def\vac{|{\rm vac}\rangle}
\def\vacn{{|{\rm vac}\rangle}_{\rm NS}}
\def\vacr{{|{\rm vac}\rangle}_{\rm R}}
\def\dvac{\langle{\rm vac}|}
\def\dvacn{{}_{\rm NS}{\langle{\rm vac}|}}
\def\dvacr{{}_{\rm R}{\langle{\rm vac}|}}
\def\br#1{\langle #1 \rangle}
\def\bra#1{\langle #1 |}
\def\ket#1{| #1 \rangle}
\def\prn{\Phi_{\rm R}^{\rm NS}}
\def\pnr{\Phi_{\rm NS}^{\rm R}}
\def\prnp{\Phi_{\rm R}^{{\rm NS}\,+}}
\def\prnm{\Phi_{\rm R}^{{\rm NS}\,-}}
\def\pnrp{\Phi^{\rm R}_{{\rm NS}\,+}}
\def\pnrm{\Phi^{\rm R}_{{\rm NS}\,-}}
\def\tr{{\rm tr}}
\def\w(#1,#2,#3,#4,#5){W\left(\matrix{#4&#3\cr#1&#2\cr}\Bigm|\,#5\right)}
\def\ww(#1,#2,#3,#4){W\left(\matrix{#4&#3\cr#1&#2\cr}\right)}
\def\wb(#1,#2,#3,#4,#5,#6)
{\overline{W}^1_#6\left(\matrix{#4&#3\cr#1&#2\cr}\Bigm|\,#5\right)}
\def\wt(#1,#2,#3,#4,#5){\widetilde{W}
\left(\matrix{#4&#3\cr#1&#2\cr}\Bigm|\,#5\right)}
\def\hom{{\rm Hom}_{U'}}
\def\goto#1{\buildrel #1 \over \longrightarrow}
\def\n{\nonumber\\}
\def\id{{\rm id}}
\def\lb{{\bf l}}
\def\Phit{\widetilde{\Phi}}
\def\Rb{\overline{R}}
\def\Remark{\medskip\noindent {\sl Remark.}\quad}
\def\vep{\varepsilon}
\def\bydef{\buildrel {\scriptstyle def} \over =}
\def\bw{\bar{w}}
\def\bW{\overline{W}}

\def\bH{{\overline \H}}
\def\Th{\Theta_p}
\def\A{{\cal A}}
\def\nnb{\nonumber}
\def\pn{\phi^{{\rm NS}}}
\def\pr{\phi^{{\rm R}}}
\def\psn{\psi^{{\rm NS}}}
\def\psr{\psi^{{\rm R}}}
\def\psz{\psi^{{\rm NS}}_0}
\def\pt{\widetilde{\phi}}
\def\pnh{\widehat{\phi}^{\rm NS}}
\def\prh{\widehat{\phi}^{\rm R}}
\def\bK{I}
\def\rN{{\rm NS}}
\def\rR{{\rm R}}
\def\rH{{\rm H}}
\def\bu{\bullet}
\def\ve{\varepsilon}
\def\refto#1{{#1}}
\def\ip(#1){(#1)_\infty}
\def\w(#1,#2,#3,#4,#5){W\left(\matrix{#4&#3\cr#1&#2\cr}\Bigm|\,#5\right)}
\def\l{\sigma}

\def\la{\lambda}
\def\slt{\goth{sl}(2)}
\def\slth{\widehat{\goth{sl}}(2)\hskip 1pt}
\def\Up{U'_q\bigl(\slth\bigr)}
\def\u{U_q\bigl(\slth \bigr)}
\def\uf{U_q\bigl(\slt\bigr)}

\section{Introduction}

In this paper, we continue our study of the role of vertex
operators and difference equations in computing
correlation functions in off-critical solvable lattice models.
We start with a review of the conformal field theory (CFT)
approach to computing critical correlation functions
in WZW-type models: models with an affine Lie algebra symmetry,
then compare it with our approach in solvable lattice models of
the vertex type.

\subsection{At criticality}
In CFT, the correlation functions on the projective plane
${\bf P}^1$:
\begin{eq}
\langle\la_0|\Phi^{\lambda_0,V^1}_{\lambda_1}(\zeta_1)
\cdots\Phi^{\lambda_{n-1},V^n}_{\lambda_n}(\zeta_n)|\la_n\rangle
\end{eq}
are given in terms of vertex operators acting as intertwiners in
the following sense:
\begin{eq}
\Phi^{\la_{j-1},V^j}_{\la_j}(\zeta):V(\la_j)\rightarrow
V(\la_{j-1})\otimes V^j_\zeta
\end{eq}
where $V(\lambda)$ is a highest weight representation with
highest weight $\lambda$ and highest weight vector $|\lambda\rangle$,
and $V^j_\zeta$ is a finite dimensional representation with a spectral
parameter $\zeta$.
For simplicity, we take all $V^j$'s to be isomorphic to
the same finite dimensional representation $V$.

The correlation functions satisfy the KZ equation, a system
of differential equations with respect to the variables
$\zeta_1,\ldots,\zeta_n$.
The solutions to the KZ equation have a branch point
at $\zeta_j=\zeta_k$, and
their monodromy is dictated by the commutation relations
of the vertex operators
\begin{eq}
\Phi^{\la_1,V^1}_{\la_4}(\zeta_1)\Phi^{\la_4,V^2}_{\la_3}(\zeta_2)
&=&\sum_{\la_2}\Phi^{\la_1,V^2}_{\la_2}(\zeta_2)\Phi^{\la_2,V^1}_{\la_3}(\zeta_1)
\ww(\la_2,\la_3,\la_4,\la_1),
\end{eq}
which is understood in terms of an appropriate analytic continuation,
and the connection coefficients $W$ are constants.

\subsection{Off criticality}
Away from criticality, one expects appropriate generalizations, or
deformations of the structures that appear at criticality.
In \cite{FR}, $q$-deformations of the
vertex operators and the KZ equation were obtained.
There is a number of major differences between the critical, or classical
(i.e. $q = 1$ ) and q-deformed cases.
To start with, the $q$-deformed KZ equation is a system of
difference rather than differential equations. Furthermore,
the commutation relation changes to
\begin{eq}
&&R^{V^1,V^2}(\zeta_1/\zeta_2)\Phi^{\la_1,V^1}_{\la_4}(\zeta_1)
\Phi^{\la_4,V^2}_{\la_3}(\zeta_2)\nonumber\\
&&=\sum_{\la_2}\Phi^{\la_1,V^2}_{\la_2}(\zeta_2)\Phi^{\la_2,V^1}_{\la_3}(\zeta_1)
\w(\la_2,\la_3,\la_4,\la_1,\zeta_1/\zeta_2).
\end{eq}
thus, the $R$-matrix appears on the left hand side,
and the connection coefficients $W$ now
depend on the spectral parameters. In the $q$-deformed case, the
multivaluedness reduces to an infinite set of poles
at $\zeta_j/\zeta_k=q^n$ for certain integers $n\ne0$.

Another crucial difference is that, in the $q$-deformed case, it
{\em matters} whether the finite dimensional
part $V^j_\zeta$ appears as the right or left component in the action
of the vertex operator: there are {\em two} different
vertex operators, depending on the position of $V^j_\zeta$
in the action. In \cite{DFJMN,JMMN},
they are distinguished as type I
($V^j_\zeta$ is to the right) and type II ($V^j_\zeta$ is to the left)
vertex operator. In this language, (2)
shows the action of a type I vertex operator.
The operators that appear in (4) are also of type I.
They are used to compute the correlation functions in the form
\begin{eq}
&&{\rm tr}_{V(\lambda_0)}
q^{-2\rho}\Phi^{\lambda_0,V^1}_{\lambda_1}(\zeta_1)
\cdots\Phi^{\lambda_{n-1},V^n}_{\lambda_n}(\zeta_n),
\end{eq}
where $\rho$ is a certain grading operator, and $\la_0=\la_n$.
\thanks{The above form of the correlation functions suggests a connection to
conformal field theory defined on the annulus.
Recently, an interesting paper appeared
in this direction \cite{Et}, where the classical and twisted
versions of the trace function were found to satisfy the classical and
elliptic versions of the KZ equation, respectively.}
On the other hand, the Type II vertex operators
serve as creation and annihilation operators that diagonalize
the Hamiltonian.

\subsection{Previous results}
In our previous work on the off-critical correlation functions
we started with the XXZ spin
chain, then considered the 8-vertex model.
In the case of the XXZ spin chain, we followed
the representation theoretic approach, since that was
possible.
The trace correlations that we obtained, as in (5),
also satisfy a version of the $q$-deformed KZ equation
(see (5) in \cite{JMN})
although the derivation in this case is different from that in
the case of correlations that are matrix
elements as in (1): it relies on the cyclic property of the trace
and the following commutation relations of the vertex operators,
which are valid in the principal grading for $V_z$:
\begin{eq}
\zeta^{\rho}\Phi^{\lambda,V}_{\lambda'}(\zeta')&=&
\Phi^{\lambda,V}_{\lambda'}(\zeta'/\zeta)\zeta^{\rho},\\
R^{V^1,V^2}(\zeta_1/\zeta_2)\Phi^{\la_1,V^1}_{\la_2}(\zeta_1)
\Phi^{\la_2,V^2}_{\la_3}(\zeta_2)
&=&\Phi^{\la_1,V^2}_{\la_2}(\zeta_2)\Phi^{\la_2,V^1}_{\la_3}(\zeta_1).
\end{eq}
Compare (4) with (7):
The reason that the summation on the right hand side disappears is that the
level of the highest weights and the finite dimensional
representation $V$ are such that
that the highest weight $\lambda$ in
\begin{eq}
\Phi^{\la,V}_{\la'}(\zeta):V(\la')\rightarrow V(\la)\otimes V_\zeta,
\end{eq}
is uniquely determined from $\lambda'$.
Specifically, in the case of the XXZ spin chain, we have:
level = 1, and $\dim V=2$ for $\slth$; refer to the
definition of {\it perfect crystals} in \cite{KKMMNN}.
This trivializes the
connection coefficients $W$, that would otherwise appear on
the right hand side.

The above simplification was used in the
derivation of the $q$-difference equations in the
the 8-vertex model (see (6) in \cite{JMN}).
In \cite{JMN}, an argument without explicit use of representation
theory, for deriving similar equations in the context of
the 8-vertex model was proposed.
The idea in this case is an extension of Baxter's corner
transfer matrix (CTM).
We replace $V(\lambda)$ with the space $\H^{(i)}$
spanned by the eigenvectors of the CTM with the specific,
say $i$-th boundary condition, $q$ is regarded as the crossing
parameter, and $\zeta^{-\rho}$ as the CTM. The basic idea is
to interpret the vertex operator as a graphical insertion of a half line.
By doing that, the commutation relations (6) and (7) follow from a
simple graphical argument, and as a corollary the elliptic difference
equation is shown to be valid for the correlation functions
of the eight-vertex model. A mathematical setting for the eight-vertex model,
in which the vertex operators are given a mathematical basis,
in terms of e.g. representation theory, is still unavailable.

\subsection{Outline of results}
Now, we turn to the subject and results of the present paper.
We wish to pursue both the graphical and the
representation theory approaches to the vertex operators and
difference equations, but this time in the context of the
ABF models.

In Section 1, we go through the graphical procedure followed in the
case of the 8-vertex model, in order to obtain the vertex operators
and the $q$-difference equations of the ABF
models. The commutation relation reads as
\begin{eq}
&&\Phi^{(i,i+1)}(\zeta_1)_{l_1,l_2}\Phi^{(i+1,i)}(\zeta_2)_{l_2,l_3}=\nonumber\\
&&\sum_{l_4}\w(l_2,l_3,l_4,l_1,\zeta_1/\zeta_2)
\Phi^{(i,i+1)}(\zeta_2)_{l_1,l_4}\Phi^{(i+1,i)}(\zeta_1)_{l_4,l_3},
\end{eq}
where the indices $i$ and $l$ label the different sectors in the
space of states, and the heights on the lattice, respectively
as will be explained later.
Next, using the results of \cite{JMOh}, we give a representation theoretic
realization of the vertex operators satisfying the above commutation
relations.

At this point, let us emphasize the following:
The graphical derivation of the difference
equations applies to other solvable models, even
when the representation theoretic setting is not
available. Examples are edge interaction models, such
as the Kashiwara-Miwa and Chiral Potts models.
We give a brief derivation of the difference
equations of the Kashiwara-Miwa model. The treatment
is basically a special case of that of the ABF models.

In the case of the XXZ spin chain, we have, not only the
representation theoretic realization of
the vertex operators, but also a more powerful realization in terms of
free bosons. In \cite{JMMN}, an integral representation was obtained
for the correlation functions using bosonization. In the ABF case,
we do not know of a bosonization, or any other tool, that
would provide us with an integral formula for the correlation functions.
However, in the case of the Ising model,
which is an intersection of all the models discussed above,
with the exception of the XXZ spin chain, an alternative tool
is available: the model can be formulated in terms of free
fermions.

The diagonalization of the Ising CTM was given in
\cite{BaxCTM} and \cite{Dav} using Jordan-Wigner fermions.
In this case, the role of type II vertex operators is basically
played by
the diagonalized fermions. In Section 2, we use this formalism
to obtain all the matrix elements of vertex operators, and
solve the difference equation for the simplest case to obtain
the two point trace function.

\section{The ABF models}

In this section, after reviewing the corner transfer matrix (CTM)
method, we introduce the
vertex operators (VO's) of the ABF models.
We deduce the commutation relations among them, express
the correlation functions as traces of VO's, and derive $q$-difference
equations for the traces.
The arguments rely on the assumption that the relevant operators
are well defined in the infinite lattice limit.
In the last subsection we relate this heuristic construction to
the representation theoretical formulation proposed earlier in
\cite{JMOh}, showing that the latter provides a mathematical model
for the VO's satisfying all the expected properties.

\subsection{Definitions}

We consider the ABF models on a square lattice [1],
and associate to each site $j$ a local state variable
$\l_j \in l_j=1,\ldots,L-1$, where $L$ is a positive integer ($\ge
4$). Next, we draw oriented lines on the dual lattice,
and associate a spectral parameter
$\zeta_H$ with each horizontal line and $\zeta_V$ with each vertical line:

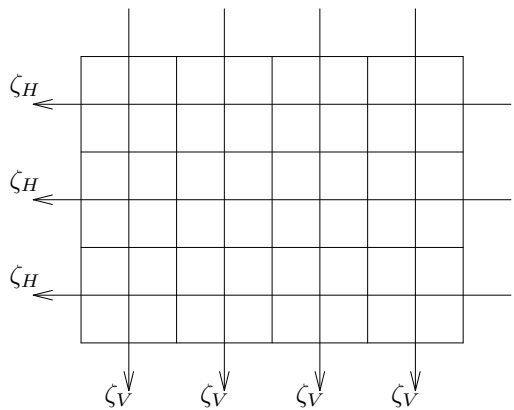
\setlength{\unitlength}{0.0125in}
\begin{picture}(214,180)(0,-10)
\drawline(210,45)(10,45)
\drawline(18.000,47.000)(10.000,45.000)(18.000,43.000)
\drawline(210,85)(10,85)
\drawline(18.000,87.000)(10.000,85.000)(18.000,83.000)
\drawline(210,125)(10,125)
\drawline(18.000,127.000)(10.000,125.000)(18.000,123.000)
\drawline(170,165)(170,5)
\drawline(168.000,13.000)(170.000,5.000)(172.000,13.000)
\drawline(130,165)(130,5)
\drawline(128.000,13.000)(130.000,5.000)(132.000,13.000)
\drawline(90,165)(90,5)
\drawline(88.000,13.000)(90.000,5.000)(92.000,13.000)
\drawline(50,165)(50,5)
\drawline(48.000,13.000)(50.000,5.000)(52.000,13.000)
\drawline(30,65)(190,65)
\drawline(30,105)(190,105)
\drawline(150,145)(150,25)
\drawline(110,145)(110,25)
\drawline(70,145)(70,25)
\drawline(30,145)(190,145)(190,25)
	(30,25)(30,145)
\put(160,0){\makebox(0,0)[lb]{\raisebox{0pt}[0pt][0pt]{\shortstack[l]{{\twlrm
$\zeta_V$}}}}}
\put(120,0){\makebox(0,0)[lb]{\raisebox{0pt}[0pt][0pt]{\shortstack[l]{{\twlrm
$\zeta_V$}}}}}
\put(80,0){\makebox(0,0)[lb]{\raisebox{0pt}[0pt][0pt]{\shortstack[l]{{\twlrm
$\zeta_V$}}}}}
\put(40,0){\makebox(0,0)[lb]{\raisebox{0pt}[0pt][0pt]{\shortstack[l]{{\twlrm
$\zeta_V$}}}}}
\put(0,50){\makebox(0,0)[lb]{\raisebox{0pt}[0pt][0pt]{\shortstack[l]{{\twlrm
$\zeta_H$}}}}}
\put(0,90){\makebox(0,0)[lb]{\raisebox{0pt}[0pt][0pt]{\shortstack[l]{{\twlrm
$\zeta_H$}}}}}
\put(0,130){\makebox(0,0)[lb]{\raisebox{0pt}[0pt][0pt]{\shortstack[l]{{\twlrm
$\zeta_H$}}}}}
\end{picture}
\noindent

To each configuration of local variables $(\l_1,\l_2,\l_3,\l_4)$
surrounding a face, we associate a Boltzmann weight. The latter
will be non-vanishing only if the configuration is admissible:
if the local sites $j$ and $j'$ are connected by an edge
then $\l_j-\l_{j'}=\pm1$. The Boltzmann weights depend on
$\zeta=\zeta_V/\zeta_H$, and also on a parameter $x$ such that
$0<x<1$, which can be thought of as temperature variable:
$x \rightarrow 0$ corresponds to the low temperature limit;
$x \rightarrow 1$ corresponds to the critical limit.
We shall restrict our attention to the region
\[
0<x<\zeta<1
\]
which corresponds to regime III in \cite{ABF}.
Since $x$ will remain constant, we will display
only the dependence on $\zeta$ and suppress the dependence on $x$;
for instance, the CTM below will be written as $A^{(i)}_4(\zeta)$.

Set $p=x^{2L}$, and define
\begin{eq}
&&\Th(z)=(z;p)_\infty(p/z;p)_\infty(p;p)_\infty, \n
&&(z;q_1,\ldots,q_k)_\infty=
\prod_{n_1,\ldots,n_k=0}^\infty(1-zq_1^{n_1}\cdots q_k^{n_k}).\nonumber
\end{eq}
With these notations, the Boltzmann weights
for the admissible configurations are

\setlength{\unitlength}{0.0125in}
\begin{picture}(114,126)(0,-10)
\drawline(31,87)(80,87)(80,37)
	(31,37)(31,87)
\drawline(56,111)(56,13)
\drawline(54.000,21.000)(56.000,13.000)(58.000,21.000)
\drawline(105,62)(6,62)
\drawline(14.000,64.000)(6.000,62.000)(14.000,60.000)
\put(0,71){\makebox(0,0)[lb]{\raisebox{0pt}[0pt][0pt]{\shortstack[l]{{\twlrm
$\zeta_H$}}}}}
\put(43,0){\makebox(0,0)[lb]{\raisebox{0pt}[0pt][0pt]{\shortstack[l]{{\twlrm
$\zeta_V$}}}}}
\put(10,20){\makebox(0,0)[lb]{\raisebox{0pt}[0pt][0pt]{\shortstack[l]{{\twlrm
$\l_1=l_1$}}}}}
\put(75,20){\makebox(0,0)[lb]{\raisebox{0pt}[0pt][0pt]{\shortstack[l]{{\twlrm
$\l_2=l_2$}}}}}
\put(75,95){\makebox(0,0)[lb]{\raisebox{0pt}[0pt][0pt]{\shortstack[l]{{\twlrm
$\l_3=l_3$}}}}}
\put(10,95){\makebox(0,0)[lb]{\raisebox{0pt}[0pt][0pt]{\shortstack[l]{{\twlrm
$\l_4=l_4$}}}}}
\put(114,59){\makebox(0,0)[lb]{\raisebox{0pt}[0pt][0pt]{\shortstack[l]{{\twlrm
$=\w(l_1,l_2,l_3,l_4,\zeta)\quad(\zeta=\zeta_V/\zeta_H)$}}}}}
\end{picture}
\begin{eq}
\w(l\pm1,l\pm2,l\pm1,l,\zeta)
&=&{1\over\kappa(\zeta)}\zeta{\Th(x^2\zeta^{-2})\over\Th(x^2)},
\label{eqn:bol1}\\
\w(l\pm1,l,l\pm1,l,\zeta)&=&{1\over\kappa(\zeta)}
{\Th(x^{2l}\zeta^{\pm2})\over\Th(x^{2l})},\label{eqn:bol2}\\
\w(l\mp1,l,l\pm1,l,\zeta)
&=&{1\over\kappa(\zeta)}
{x\over\zeta}{\Th(\zeta^2)\Th(x^{2(l\pm1)})\over\Th(x^2)\Th(x^{2l})}.
\label{eqn:bol3}
\end{eq}
Here the normalizing factor $\kappa(\zeta)$ is so chosen
that the partition function per site is equal to $1$.
{}From the standard inversion trick we find
\begin{eq}
\kappa(\zeta)&=&{\bar\kappa}(\zeta){\Th(x^2\zeta^{-2})\over\Th(x^2)},\\
{\bar\kappa}(\zeta)&=&
{\ip(x^2\zeta^2;p,x^4)
\ip(px^2\zeta^2;p,x^4)
\ip(x^4\zeta^{-2};p,x^4)
\ip(p\zeta^{-2};p,x^4)
\over
\ip(x^2\zeta^{-2};p,x^4)
\ip(px^2\zeta^{-2};p,x^4)
\ip(x^4\zeta^2;p,x^4)
\ip(p\zeta^2;p,x^4)} \label{eqn:kappa}
\end{eq}
Note that $\kappa(\zeta)=\kappa(x\zeta^{-1})$.
For later use we set
\begin{eq}
g_l&=&\Th(x^{2l}).\label{eqn:gl}
\end{eq}

Instead of taking the $\zeta_H, \zeta_V$ to be all the same, we could equally
well assign spectral parameters that can vary independently from line to line.
The resulting model is  $Z$-invariant in Baxter's sense.
Later we will take advantage of this freedom to vary some of the spectral
parameters.

The basic properties of the Boltzmann weights are as follows:

\medskip
\noindent{\bf Initial condition}
\begin{eq}
\w(l_1,l_2,l_3,l_4,1)&=&\delta_{l_1,l_3}.
\label{eqn:init}
\end{eq}

\noindent{\bf Unitarity relation}
\begin{eq}
\sum_{l}\w(l_1,l_2,l,l_4,\zeta^{-1})\w(l,l_2,l_3,l_4,\zeta)&=&
\delta_{l_1,l_3}.
\label{eqn:uni}
\end{eq}

\setlength{\unitlength}{0.0125in}
\begin{picture}(150,139)(0,-10)
\drawline(35,70)(75,70)(75,30)
	(35,30)(35,70)
\drawline(75,110)(115,110)(115,70)
	(75,70)(75,110)
\drawline(135,90)	(131.959,90.363)
	(129.020,90.703)
	(126.181,91.020)
	(123.439,91.312)
	(120.792,91.582)
	(118.237,91.828)
	(113.391,92.250)
	(108.878,92.578)
	(104.678,92.812)
	(100.767,92.953)
	(97.125,93.000)
	(93.729,92.953)
	(90.557,92.812)
	(87.587,92.578)
	(84.797,92.250)
	(82.165,91.828)
	(79.670,91.312)
	(75.000,90.000)

\drawline(75,90)	(72.254,88.895)
	(69.344,87.156)
	(66.387,84.902)
	(63.500,82.250)
	(60.801,79.316)
	(58.406,76.219)
	(56.434,73.074)
	(55.000,70.000)

\drawline(55,70)	(53.688,65.658)
	(52.750,60.766)
	(52.422,58.058)
	(52.188,55.146)
	(52.047,52.010)
	(52.000,48.625)
	(52.047,44.971)
	(52.188,41.025)
	(52.422,36.766)
	(52.750,32.172)
	(53.172,27.220)
	(53.418,24.603)
	(53.688,21.889)
	(53.980,19.074)
	(54.297,16.156)
	(54.637,13.132)
	(55.000,10.000)

\drawline(52.079,17.712)(55.000,10.000)(56.052,18.179)
\drawline(95,124)	(95.422,120.887)
	(95.816,117.881)
	(96.184,114.980)
	(96.524,112.181)
	(96.837,109.481)
	(97.122,106.877)
	(97.612,101.947)
	(97.993,97.370)
	(98.265,93.124)
	(98.429,89.186)
	(98.483,85.534)
	(98.429,82.147)
	(98.265,79.003)
	(97.993,76.079)
	(97.612,73.354)
	(96.524,68.411)
	(95.000,64.000)

\drawline(95,64)	(93.592,61.237)
	(91.635,58.313)
	(89.247,55.345)
	(86.548,52.452)
	(83.655,49.753)
	(80.687,47.365)
	(77.763,45.408)
	(75.000,44.000)

\drawline(75,44)	(70.589,42.476)
	(65.646,41.388)
	(62.921,41.007)
	(59.997,40.735)
	(56.853,40.571)
	(53.466,40.517)
	(49.814,40.571)
	(45.876,40.735)
	(41.630,41.007)
	(37.053,41.388)
	(32.123,41.878)
	(29.519,42.163)
	(26.819,42.476)
	(24.020,42.816)
	(21.119,43.184)
	(18.113,43.578)
	(15.000,44.000)

\drawline(23.198,44.892)(15.000,44.000)(22.653,40.929)
\put(80,60){\makebox(0,0)[lb]{\raisebox{0pt}[0pt][0pt]{\shortstack[l]{{\twlrm
$l$}}}}}
\put(25,70){\makebox(0,0)[lb]{\raisebox{0pt}[0pt][0pt]{\shortstack[l]{{\twlrm
$l_4$}}}}}
\put(60,110){\makebox(0,0)[lb]{\raisebox{0pt}[0pt][0pt]{\shortstack[l]{{\twlrm
$l_4$}}}}}
\put(120,110){\makebox(0,0)[lb]{\raisebox{0pt}[0pt][0pt]{\shortstack[l]{{\twlrm
$l_3$}}}}}
\put(120,60){\makebox(0,0)[lb]{\raisebox{0pt}[0pt][0pt]{\shortstack[l]{{\twlrm
$l_2$}}}}}
\put(140,60){\makebox(0,0)[lb]{\raisebox{0pt}[0pt][0pt]{\shortstack[l]{{\twlrm
$=\quad\delta_{l_1,l_3}$}}}}}
\put(80,20){\makebox(0,0)[lb]{\raisebox{0pt}[0pt][0pt]{\shortstack[l]{{\twlrm
$l_2$}}}}}
\put(25,20){\makebox(0,0)[lb]{\raisebox{0pt}[0pt][0pt]{\shortstack[l]{{\twlrm
$l_1$}}}}}
\put(45,0){\makebox(0,0)[lb]{\raisebox{0pt}[0pt][0pt]{\shortstack[l]{{\twlrm
$\zeta_H$}}}}}
\put(0,40){\makebox(0,0)[lb]{\raisebox{0pt}[0pt][0pt]{\shortstack[l]{{\twlrm
$\zeta_V$}}}}}
\end{picture}
\noindent{\bf Crossing symmetry}
\begin{eq}
&&\w(l_4,l_1,l_2,l_3,\zeta^{-1})=\w(l_1,l_2,l_3,l_4,x\zeta)
{g_{l_2}\over g_{l_3}}
\label{eqn:cross}
\end{eq}

\setlength{\unitlength}{0.0125in}
\begin{picture}(254,110)(0,-10)
\drawline(90,55)(10,55)
\drawline(18.000,57.000)(10.000,55.000)(18.000,53.000)
\drawline(30,75)(70,75)(70,35)
	(30,35)(30,75)
\drawline(50,15)(50,95)
\drawline(52.000,87.000)(50.000,95.000)(48.000,87.000)
\drawline(155,75)(195,75)(195,35)
	(155,35)(155,75)
\drawline(175,95)(175,15)
\drawline(173.000,23.000)(175.000,15.000)(177.000,23.000)
\drawline(215,55)(135,55)
\drawline(143.000,57.000)(135.000,55.000)(143.000,53.000)
\put(222,52){\makebox(0,0)[lb]{\raisebox{0pt}[0pt][0pt]{\shortstack[l]{{\twlrm
$\displaystyle g_{l_2}\over\displaystyle g_{l_3}$,}}}}}
\put(125,60){\makebox(0,0)[lb]{\raisebox{0pt}[0pt][0pt]{\shortstack[l]{{\twlrm
$\zeta_H$}}}}}
\put(165,0){\makebox(0,0)[lb]{\raisebox{0pt}[0pt][0pt]{\shortstack[l]{{\twlrm
$x\zeta_V$}}}}}
\put(40,0){\makebox(0,0)[lb]{\raisebox{0pt}[0pt][0pt]{\shortstack[l]{{\twlrm
$\zeta_V$}}}}}
\put(0,60){\makebox(0,0)[lb]{\raisebox{0pt}[0pt][0pt]{\shortstack[l]{{\twlrm
$\zeta_H$}}}}}
\put(20,80){\makebox(0,0)[lb]{\raisebox{0pt}[0pt][0pt]{\shortstack[l]{{\twlrm
$l_4$}}}}}
\put(70,80){\makebox(0,0)[lb]{\raisebox{0pt}[0pt][0pt]{\shortstack[l]{{\twlrm
$l_3$}}}}}
\put(70,25){\makebox(0,0)[lb]{\raisebox{0pt}[0pt][0pt]{\shortstack[l]{{\twlrm
$l_2$}}}}}
\put(15,25){\makebox(0,0)[lb]{\raisebox{0pt}[0pt][0pt]{\shortstack[l]{{\twlrm
$l_1$}}}}}
\put(135,80){\makebox(0,0)[lb]{\raisebox{0pt}[0pt][0pt]{\shortstack[l]{{\twlrm
$l_4$}}}}}
\put(200,80){\makebox(0,0)[lb]{\raisebox{0pt}[0pt][0pt]{\shortstack[l]{{\twlrm
$l_3$}}}}}
\put(200,25){\makebox(0,0)[lb]{\raisebox{0pt}[0pt][0pt]{\shortstack[l]{{\twlrm
$l_2$}}}}}
\put(140,25){\makebox(0,0)[lb]{\raisebox{0pt}[0pt][0pt]{\shortstack[l]{{\twlrm
$l_1$}}}}}
\put(105,55){\makebox(0,0)[lb]{\raisebox{0pt}[0pt][0pt]{\shortstack[l]{{\twlrm
$=$}}}}}
\end{picture}
\vskip-10pt
\noindent{\bf Yang-Baxter equation}
\begin{eq}
&&\sum_l \w(l,l_2,l_3,l_4,\zeta_2/\zeta_3)
\w(l_6,l,l_4,l_5,\zeta_1/\zeta_3)\w(l_1,l_2,l,l_6,\zeta_1/\zeta_2) \n
&&=
\sum_l \w(l,l_3,l_4,l_5,\zeta_1/\zeta_2)
\w(l_1,l_2,l_3,l,\zeta_1/\zeta_3)\w(l_6,l_1,l,l_5,\zeta_2/\zeta_3).
\label{eqn:YBE}
\end{eq}

\setlength{\unitlength}{0.0125in}
\begin{picture}(315,180)(0,-10)
\drawline(295,45)(175,45)
\drawline(183.000,47.000)(175.000,45.000)(183.000,43.000)
\drawline(235,65)(235,25)
\drawline(195,65)(275,65)(275,25)
	(195,25)(195,65)
\drawline(235,145)(195,105)(235,65)
	(275,105)(235,145)
\drawline(135,125)(15,125)
\drawline(23.000,127.000)(15.000,125.000)(23.000,123.000)
\drawline(75,105)(35,65)(75,25)
	(115,65)(75,105)
\drawline(75,145)(75,105)
\drawline(35,145)(115,145)(115,105)
	(35,105)(35,145)
\drawline(195,145)	(198.535,142.309)
	(201.935,139.698)
	(205.204,137.164)
	(208.343,134.705)
	(211.356,132.317)
	(214.246,129.999)
	(217.014,127.748)
	(219.665,125.560)
	(224.623,121.363)
	(229.142,117.389)
	(233.243,113.614)
	(236.949,110.017)
	(240.282,106.576)
	(243.263,103.268)
	(245.914,100.072)
	(248.259,96.966)
	(250.317,93.928)
	(252.112,90.935)
	(253.666,87.967)
	(255.000,85.000)

\drawline(255,85)	(256.088,82.133)
	(257.032,79.118)
	(257.830,75.927)
	(258.483,72.529)
	(258.991,68.895)
	(259.354,64.997)
	(259.571,60.804)
	(259.644,56.288)
	(259.571,51.419)
	(259.481,48.844)
	(259.354,46.169)
	(259.190,43.391)
	(258.991,40.507)
	(258.755,37.512)
	(258.483,34.404)
	(258.175,31.178)
	(257.830,27.831)
	(257.449,24.359)
	(257.032,20.759)
	(256.578,17.027)
	(256.088,13.158)
	(255.562,9.151)
	(255.000,5.000)

\drawline(254.108,13.198)(255.000,5.000)(258.071,12.653)
\drawline(275,145)	(271.465,142.309)
	(268.065,139.698)
	(264.796,137.164)
	(261.657,134.705)
	(258.644,132.317)
	(255.754,129.999)
	(252.986,127.748)
	(250.335,125.560)
	(245.377,121.363)
	(240.858,117.389)
	(236.757,113.614)
	(233.051,110.017)
	(229.718,106.576)
	(226.737,103.268)
	(224.086,100.072)
	(221.741,96.966)
	(219.683,93.928)
	(217.888,90.935)
	(216.334,87.967)
	(215.000,85.000)

\drawline(215,85)	(213.912,82.133)
	(212.968,79.118)
	(212.170,75.927)
	(211.517,72.529)
	(211.009,68.895)
	(210.646,64.997)
	(210.429,60.804)
	(210.356,56.288)
	(210.429,51.419)
	(210.519,48.844)
	(210.646,46.169)
	(210.810,43.391)
	(211.009,40.507)
	(211.245,37.512)
	(211.517,34.404)
	(211.825,31.178)
	(212.170,27.831)
	(212.551,24.359)
	(212.968,20.759)
	(213.422,17.027)
	(213.912,13.158)
	(214.438,9.151)
	(215.000,5.000)

\drawline(211.929,12.653)(215.000,5.000)(215.892,13.198)
\drawline(95,165)	(95.562,160.849)
	(96.088,156.842)
	(96.578,152.973)
	(97.032,149.241)
	(97.449,145.641)
	(97.830,142.169)
	(98.175,138.822)
	(98.483,135.596)
	(98.755,132.488)
	(98.991,129.493)
	(99.190,126.609)
	(99.354,123.831)
	(99.481,121.156)
	(99.571,118.581)
	(99.644,113.712)
	(99.571,109.196)
	(99.354,105.003)
	(98.991,101.105)
	(98.483,97.471)
	(97.830,94.073)
	(97.032,90.882)
	(96.088,87.867)
	(95.000,85.000)

\drawline(95,85)	(93.666,82.033)
	(92.112,79.065)
	(90.317,76.072)
	(88.259,73.034)
	(85.914,69.928)
	(83.263,66.732)
	(80.282,63.424)
	(76.949,59.983)
	(73.243,56.386)
	(69.142,52.611)
	(64.623,48.637)
	(59.665,44.440)
	(57.014,42.252)
	(54.246,40.001)
	(51.356,37.683)
	(48.343,35.295)
	(45.204,32.836)
	(41.935,30.302)
	(38.535,27.691)
	(35.000,25.000)

\drawline(40.166,31.428)(35.000,25.000)(42.583,28.240)
\drawline(55,165)	(54.438,160.849)
	(53.912,156.842)
	(53.422,152.973)
	(52.968,149.241)
	(52.551,145.641)
	(52.170,142.169)
	(51.825,138.822)
	(51.517,135.596)
	(51.245,132.488)
	(51.009,129.493)
	(50.810,126.609)
	(50.646,123.831)
	(50.519,121.156)
	(50.429,118.581)
	(50.356,113.712)
	(50.429,109.196)
	(50.646,105.003)
	(51.009,101.105)
	(51.517,97.471)
	(52.170,94.073)
	(52.968,90.882)
	(53.912,87.867)
	(55.000,85.000)

\drawline(55,85)	(56.334,82.033)
	(57.888,79.065)
	(59.683,76.072)
	(61.741,73.034)
	(64.086,69.928)
	(66.737,66.732)
	(69.718,63.424)
	(73.051,59.983)
	(76.757,56.386)
	(80.858,52.611)
	(85.377,48.637)
	(90.335,44.440)
	(92.986,42.252)
	(95.754,40.001)
	(98.644,37.683)
	(101.657,35.295)
	(104.796,32.836)
	(108.065,30.302)
	(111.465,27.691)
	(115.000,25.000)

\drawline(107.417,28.240)(115.000,25.000)(109.834,31.428)
\put(235,75){\makebox(0,0)[lb]{\raisebox{0pt}[0pt][0pt]{\shortstack[l]{{\twlrm
$l$}}}}}
\put(75,90){\makebox(0,0)[lb]{\raisebox{0pt}[0pt][0pt]{\shortstack[l]{{\twlrm
$l$}}}}}
\put(25,100){\makebox(0,0)[lb]{\raisebox{0pt}[0pt][0pt]{\shortstack[l]{{\twlrm
$l_6$}}}}}
\put(25,60){\makebox(0,0)[lb]{\raisebox{0pt}[0pt][0pt]{\shortstack[l]{{\twlrm
$l_6$}}}}}
\put(30,150){\makebox(0,0)[lb]{\raisebox{0pt}[0pt][0pt]{\shortstack[l]{{\twlrm
$l_5$}}}}}
\put(70,150){\makebox(0,0)[lb]{\raisebox{0pt}[0pt][0pt]{\shortstack[l]{{\twlrm
$l_4$}}}}}
\put(120,100){\makebox(0,0)[lb]{\raisebox{0pt}[0pt][0pt]{\shortstack[l]{{\twlrm
$l_2$}}}}}
\put(120,145){\makebox(0,0)[lb]{\raisebox{0pt}[0pt][0pt]{\shortstack[l]{{\twlrm
$l_3$}}}}}
\put(120,60){\makebox(0,0)[lb]{\raisebox{0pt}[0pt][0pt]{\shortstack[l]{{\twlrm
$l_2$}}}}}
\put(70,10){\makebox(0,0)[lb]{\raisebox{0pt}[0pt][0pt]{\shortstack[l]{{\twlrm
$l_1$}}}}}
\put(185,15){\makebox(0,0)[lb]{\raisebox{0pt}[0pt][0pt]{\shortstack[l]{{\twlrm
$l_6$}}}}}
\put(185,65){\makebox(0,0)[lb]{\raisebox{0pt}[0pt][0pt]{\shortstack[l]{{\twlrm
$l_5$}}}}}
\put(180,100){\makebox(0,0)[lb]{\raisebox{0pt}[0pt][0pt]{\shortstack[l]{{\twlrm
$l_5$}}}}}
\put(225,150){\makebox(0,0)[lb]{\raisebox{0pt}[0pt][0pt]{\shortstack[l]{{\twlrm
$l_4$}}}}}
\put(280,100){\makebox(0,0)[lb]{\raisebox{0pt}[0pt][0pt]{\shortstack[l]{{\twlrm
$l_3$}}}}}
\put(285,65){\makebox(0,0)[lb]{\raisebox{0pt}[0pt][0pt]{\shortstack[l]{{\twlrm
$l_3$}}}}}
\put(275,15){\makebox(0,0)[lb]{\raisebox{0pt}[0pt][0pt]{\shortstack[l]{{\twlrm
$l_2$}}}}}
\put(230,15){\makebox(0,0)[lb]{\raisebox{0pt}[0pt][0pt]{\shortstack[l]{{\twlrm
$l_1$}}}}}
\put(260,0){\makebox(0,0)[lb]{\raisebox{0pt}[0pt][0pt]{\shortstack[l]{{\twlrm
$\zeta_1$}}}}}
\put(205,0){\makebox(0,0)[lb]{\raisebox{0pt}[0pt][0pt]{\shortstack[l]{{\twlrm
$\zeta_2$}}}}}
\put(165,45){\makebox(0,0)[lb]{\raisebox{0pt}[0pt][0pt]{\shortstack[l]{{\twlrm
$\zeta_3$}}}}}
\put(120,20){\makebox(0,0)[lb]{\raisebox{0pt}[0pt][0pt]{\shortstack[l]{{\twlrm
$\zeta_1$}}}}}
\put(20,20){\makebox(0,0)[lb]{\raisebox{0pt}[0pt][0pt]{\shortstack[l]{{\twlrm
$\zeta_2$}}}}}
\put(0,125){\makebox(0,0)[lb]{\raisebox{0pt}[0pt][0pt]{\shortstack[l]{{\twlrm
$\zeta_3$}}}}}
\end{picture}
\subsection{Corner transfer matrices}

Let us label the vertices on the lattice as $(j_1,j_2)\ (j_1,j_2\in\Z)$
so that $j_1$ is increasing to the left and $j_2$ is increasing in the
upward direction.
For $i=0,1\in\Z/2\Z$ and $m=1,\ldots,L-2$,
we denote by $C^{(i)}_{m,m+1}$ the ground state configuration such that
\begin{eq}
\l_{(j_1,j_2)}&=&m+1\hbox{ if $j_1+j_2=i+1 \bmod2$,}\label{eqn:b.c.1}\\
&=&m\hbox{ if $j_1+j_2=i \bmod2$.}\label{eqn:b.c.2}
\end{eq}
In the following discussion we choose
and fix $m$, and suppress it in our notation.

Consider the NW-quadrant of the lattice consisting of $M+1$ rows and
$M+1$ columns, with the vertex $(1,1)$ at the SE corner:

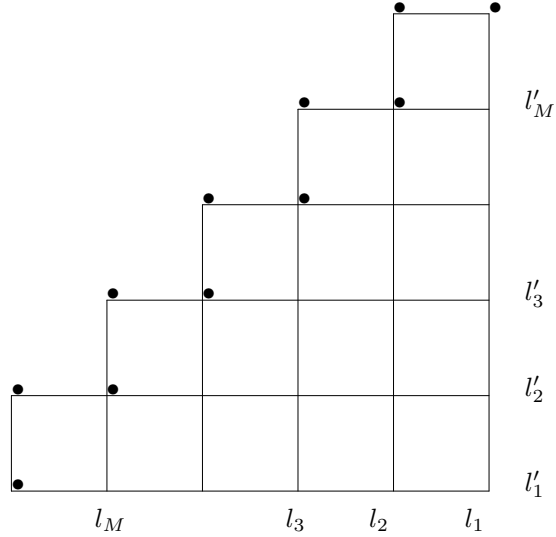
\setlength{\unitlength}{0.0125in}
\begin{picture}(254,240)(0,-10)
\drawline(200,95)(40,95)(40,15)
\drawline(200,135)(80,135)(80,15)
\drawline(200,175)(120,175)(120,15)
\drawline(200,215)(160,215)(160,15)
\drawline(200,215)(200,15)(0,15)
	(0,55)(200,55)
\put(160,175){\makebox(0,0)[lb]{\raisebox{0pt}[0pt][0pt]{\shortstack[l]{{\twlrm
$\bullet$}}}}}
\put(120,175){\makebox(0,0)[lb]{\raisebox{0pt}[0pt][0pt]{\shortstack[l]{{\twlrm
$\bullet$}}}}}
\put(120,135){\makebox(0,0)[lb]{\raisebox{0pt}[0pt][0pt]{\shortstack[l]{{\twlrm
$\bullet$}}}}}
\put(80,135){\makebox(0,0)[lb]{\raisebox{0pt}[0pt][0pt]{\shortstack[l]{{\twlrm
$\bullet$}}}}}
\put(80,95){\makebox(0,0)[lb]{\raisebox{0pt}[0pt][0pt]{\shortstack[l]{{\twlrm
$\bullet$}}}}}
\put(40,95){\makebox(0,0)[lb]{\raisebox{0pt}[0pt][0pt]{\shortstack[l]{{\twlrm
$\bullet$}}}}}
\put(40,55){\makebox(0,0)[lb]{\raisebox{0pt}[0pt][0pt]{\shortstack[l]{{\twlrm
$\bullet$}}}}}
\put(0,55){\makebox(0,0)[lb]{\raisebox{0pt}[0pt][0pt]{\shortstack[l]{{\twlrm
$\bullet$}}}}}
\put(200,215){\makebox(0,0)[lb]{\raisebox{0pt}[0pt][0pt]{\shortstack[l]{{\twlrm
$\bullet$}}}}}
\put(160,215){\makebox(0,0)[lb]{\raisebox{0pt}[0pt][0pt]{\shortstack[l]{{\twlrm
$\bullet$}}}}}
\put(0,15){\makebox(0,0)[lb]{\raisebox{0pt}[0pt][0pt]{\shortstack[l]{{\twlrm
$\bullet$}}}}}
\put(35,0){\makebox(0,0)[lb]{\raisebox{0pt}[0pt][0pt]{\shortstack[l]{{\twlrm
$l_M$}}}}}
\put(115,0){\makebox(0,0)[lb]{\raisebox{0pt}[0pt][0pt]{\shortstack[l]{{\twlrm
$l_3$}}}}}
\put(150,0){\makebox(0,0)[lb]{\raisebox{0pt}[0pt][0pt]{\shortstack[l]{{\twlrm
$l_2$}}}}}
\put(190,0){\makebox(0,0)[lb]{\raisebox{0pt}[0pt][0pt]{\shortstack[l]{{\twlrm
$l_1$}}}}}
\put(215,175){\makebox(0,0)[lb]{\raisebox{0pt}[0pt][0pt]{\shortstack[l]{{\twlrm
$l'_M$}}}}}
\put(215,95){\makebox(0,0)[lb]{\raisebox{0pt}[0pt][0pt]{\shortstack[l]{{\twlrm
$l'_3$}}}}}
\put(215,55){\makebox(0,0)[lb]{\raisebox{0pt}[0pt][0pt]{\shortstack[l]{{\twlrm
$l'_2$}}}}}
\put(215,15){\makebox(0,0)[lb]{\raisebox{0pt}[0pt][0pt]{\shortstack[l]{{\twlrm
$l'_1$}}}}}
\end{picture}

\noindent
On the boundary shown by bullets, the variables are fixed to
have the same value as in $C^{(i)}_{m,m+1}$.
Let us choose a configuration of the local variables
$\lb'=(l'_1,\ldots,l'_M)$ (resp. $\lb=(l_1,\ldots,l_M)$)
on the vertical (resp. horizontal) half line
$(1,1),\ldots,(1,M)$ (resp. $(1,1),\ldots$, $(M,1)$).
Calculate the partition function of this quadrant
while fixing these variables.
The $(\lb,\lb')$ matrix element of the CTM $A^{(i)}_4(\zeta)$
is defined to be this partition function.
Here $i$ refers to the choice of the boundary configuration.
The matrix element is defined to be zero unless $l_1=l'_1$.
It is known that in the limit $M\rightarrow\infty$
the CTM takes the simple form
\begin{eq}
&&A^{(i)}_4(\zeta)\sim\zeta^{D^{(i)}}, \nonumber
\end{eq}
where $D^{(i)}$ is the CTM Hamiltonian: a $\zeta$-independent operator
whose spectrum is contained in the set $\{0,1,2,\ldots\}$.
The symbol $\sim$ indicates equality up to multiplication by a scalar.
Such a scalar factor is irrelevant in the computation of the correlation
functions.

Set
\begin{eq}
{\bar l}^{(i)}_k&=&m\hbox{ if $k=i+1 \bmod2$,}\nonumber\\
&=&m+1\hbox{ if $k=i  \bmod2$.}\nonumber
\end{eq}
Consider the vector space $\H^{(i)}$ whose basis elements are
admissible configurations $\lb=(l_1,l_2,\ldots)$ satisfying
the boundary condition
\begin{eq}
l_k&=&{\bar l}^{(i)}_k\hbox{ if $k>\hskip-2pt>1$.}\nonumber
\end{eq}
Let us write $\ket{\lb}$ when we wish to emphasize that it is a vector
in this space.
Formally the CTM $A^{(i)}_4(\zeta)$ or $D^{(i)}$ act on $\H^{(i)}$,
and their eigenvectors are certain infinite linear combinations of the
$\ket{\lb}$.
In our heuristic approach, we identify $\H^{(i)}$ with
the vector space spanned by these eigenvectors.

We denote by $S^{(i)}_1$, $G^{(i)}_1$ the operators
\begin{eq}
S^{(i)}_1 \ket{\lb}=l_1 \ket{\lb},
\quad G^{(i)}_1\ket{\lb}=g_{l_1}\ket{\lb},\label{eqn:opSG}
\end{eq}
where $g_l$ is defined in (\ref{eqn:gl}).
By construction $D^{(i)}$ commutes with $S^{(i)}_1$.
Hence each eigenspace $\H^{(i)}_l$ of $S^{(i)}_1$ with eigenvalue $l$
is invariant under the action of $D^{(i)}$, and we have
$\H^{(i)}=\oplus_l\H^{(i)}_l$.

The CTMs for the NE, SE and SW quadrants
$A^{(i)}_1(\zeta)$, $A^{(i)}_2(\zeta)$, $A^{(i)}_3(\zeta)$ are defined
similarly.
Let $R^{(i)}$ be the diagonal matrix acting on $\H^{(i)}$
\begin{eq}
R^{(i)}\ket{\lb}&=&\prod_{k=1}^\infty
{g_{l_k}\over g_{{\bar l}^{(i)}_k}}\ket{\lb}.\label{eqn:opR}
\end{eq}
Then the crossing symmetry (\ref{eqn:cross}) implies that
\begin{eq}
A^{(i)}_3(\zeta)&\sim&R^{(i)}\cdot(x/\zeta)^{D^{(i)}},\label{eqn:CTM3}\\
A^{(i)}_2(\zeta)&\sim&R^{(i)}\cdot \zeta^{D^{(i)}}\cdot R^{(i)-1},
\label{eqn:CTM2}\\
A^{(i)}_1(\zeta)&\sim&G^{(i)}_1\cdot (x/\zeta)^{D^{(i)}}\cdot R^{(i)-1}.
\label{eqn:CTM1}
\end{eq}
Therefore, we have
\begin{eq}
A^{(i)}_1(\zeta)A^{(i)}_2(\zeta)A^{(i)}_3(\zeta)A^{(i)}_4(\zeta)
\sim&G^{(i)}_1\cdot x^{2D^{(i)}}.
\end{eq}

\subsection{Vertex operators}

To be able to write down expressions for the
correlation functions, we need to introduce the VO's
\[
\Phi^{(i+1,i)}(\zeta):\H^{(i)}\goto{}\H^{(i+1)}.
\]
Notice the difference between the vertex operators
introduced above, and those defined in the previous section.
The point is that the VO's discussed in the previous section
are defined in the context of vertex models. The VO's that
we discuss in this section are defined in the context of
face models. We will refer to the former as VO's of the {\em vertex type}
and to the latter as VO's of the {\em face type}.
The relation between them will
be discussed further below, in the subsection "Construction
by representation theory".

Graphically the $(\lb,\lb')$ matrix element of $\Phi^{(i+1,i)}(\zeta)$
is defined to be the product of Boltzmann weights as follows

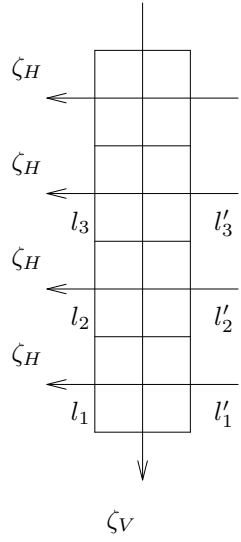
\setlength{\unitlength}{0.0125in}
\begin{picture}(121,235)(0,-10)
\drawline(55,220)(55,20)
\drawline(53.000,28.000)(55.000,20.000)(57.000,28.000)
\drawline(75,200)(75,40)(35,40)(35,200)
\drawline(75,80)(35,80)
\drawline(75,120)(35,120)
\drawline(75,160)(35,160)
\drawline(75,200)(35,200)
\drawline(95,60)(15,60)
\drawline(23.000,62.000)(15.000,60.000)(23.000,58.000)
\drawline(95,140)(15,140)
\drawline(23.000,142.000)(15.000,140.000)(23.000,138.000)
\drawline(95,100)(15,100)
\drawline(23.000,102.000)(15.000,100.000)(23.000,98.000)
\drawline(95,180)(15,180)
\drawline(23.000,182.000)(15.000,180.000)(23.000,178.000)
\put(25,45){\makebox(0,0)[lb]{\raisebox{0pt}[0pt][0pt]{\shortstack[l]{{\twlrm
$l_1$}}}}}
\put(25,85){\makebox(0,0)[lb]{\raisebox{0pt}[0pt][0pt]{\shortstack[l]{{\twlrm
$l_2$}}}}}
\put(25,125){\makebox(0,0)[lb]{\raisebox{0pt}[0pt][0pt]{\shortstack[l]{{\twlrm
$l_3$}}}}}
\put(85,125){\makebox(0,0)[lb]{\raisebox{0pt}[0pt][0pt]{\shortstack[l]{{\twlrm
$l'_3$}}}}}
\put(85,85){\makebox(0,0)[lb]{\raisebox{0pt}[0pt][0pt]{\shortstack[l]{{\twlrm
$l'_2$}}}}}
\put(85,45){\makebox(0,0)[lb]{\raisebox{0pt}[0pt][0pt]{\shortstack[l]{{\twlrm
$l'_1$}}}}}
\put(40,0){\makebox(0,0)[lb]{\raisebox{0pt}[0pt][0pt]{\shortstack[l]{{\twlrm
$\zeta_V$}}}}}
\put(0,190){\makebox(0,0)[lb]{\raisebox{0pt}[0pt][0pt]{\shortstack[l]{{\twlrm
$\zeta_H$}}}}}
\put(0,150){\makebox(0,0)[lb]{\raisebox{0pt}[0pt][0pt]{\shortstack[l]{{\twlrm
$\zeta_H$}}}}}
\put(0,110){\makebox(0,0)[lb]{\raisebox{0pt}[0pt][0pt]{\shortstack[l]{{\twlrm
$\zeta_H$}}}}}
\put(0,70){\makebox(0,0)[lb]{\raisebox{0pt}[0pt][0pt]{\shortstack[l]{{\twlrm
$\zeta_H$}}}}}
\end{picture}

The CTM's, their Hamiltonians, and the vertex operators act on
specific subspaces of the full space of states. The action on
a certain subspace $\H^{(i)}_l$ will be indicated by the
subscripts $l,l'$, as follows:
\begin{eq}
A^{(i)}_k(\zeta)_l,~D^{(i)}_l&:&\H^{(i)}_l\rightarrow\H^{(i)}_l,\nonumber\\
\Phi^{(i,i')}(\zeta)_{l,l'}&:&\H^{(i')}_{l'}\rightarrow\H^{(i)}_l.\nonumber
\end{eq}

We shall argue that,
with an appropriate choice of a scalar factor, the VO's
enjoy the following basic properties:

\medskip
\noindent{\bf Homogeneity}
\begin{eq}
&&\zeta^{D^{(i+1)}}\circ \Phi^{(i+1,i)}(\zeta')\circ \zeta^{-D^{(i)}}=
\Phi^{(i+1,i)}(\zeta'/\zeta). \label{eqn:Facehomo}
\end{eq}

\noindent{\bf Commutation relations}
\begin{eq}
&&\sum_{l_3}\w(l_1,l_2,l_3,l_4,\zeta_1/\zeta_2)
\Phi^{(i,i+1)}(\zeta_1)_{l_4,l_3}\Phi^{(i+1,i)}(\zeta_2)_{l_3,l_2}\nonumber\\
&&\hskip2cm
=\Phi^{(i,i+1)}(\zeta_2)_{l_4,l_1}\Phi^{(i+1,i)}(\zeta_1)_{l_1,l_2}.
\label{eqn:Facecomm}
\end{eq}

\noindent{\bf Normalization}
\begin{eq}
\sum_{l_2}g_{l_2}\Phi^{(i,i+1)}(x\zeta)_{l_1,l_2}
\Phi^{(i+1,i)}(\zeta)_{l_2,l_1}&=&id_{\H^{(i)}_{l_1}}.
\label{eqn:Facedual}
\end{eq}

To deduce (\ref{eqn:Facehomo}), consider the following:

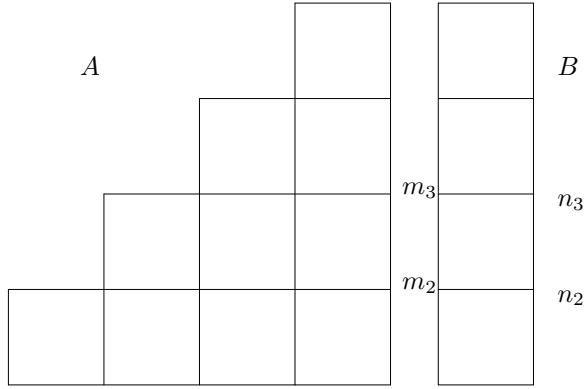
\setlength{\unitlength}{0.0125in}
\begin{picture}(260,195)(0,-10)
\drawline(180,140)(220,140)
\drawline(180,100)(220,100)
\drawline(180,60)(220,60)
\drawline(0,20)(120,20)
\drawline(160,60)(0,60)(0,20)
\drawline(160,100)(40,100)(40,20)
\drawline(160,140)(80,140)(80,20)
\drawline(120,180)(160,180)(160,20)
	(120,20)(120,180)
\drawline(180,180)(220,180)(220,20)
	(180,20)(180,180)
\put(230,0){\makebox(0,0)[lb]{\raisebox{0pt}[0pt][0pt]{\shortstack[l]{{\twlrm
$l_1=n_1$}}}}}
\put(30,150){\makebox(0,0)[lb]{\raisebox{0pt}[0pt][0pt]{\shortstack[l]{{\twlrm
$A$}}}}}
\put(230,150){\makebox(0,0)[lb]{\raisebox{0pt}[0pt][0pt]{\shortstack[l]{{\twlrm
$B$}}}}}
\put(70,0){\makebox(0,0)[lb]{\raisebox{0pt}[0pt][0pt]{\shortstack[l]{{\twlrm
$l_4=p_3$}}}}}
\put(110,0){\makebox(0,0)[lb]{\raisebox{0pt}[0pt][0pt]{\shortstack[l]{{\twlrm
$l_3=p_2$}}}}}
\put(165,100){\makebox(0,0)[lb]{\raisebox{0pt}[0pt][0pt]{\shortstack[l]{{\twlrm
$m_3$}}}}}
\put(165,60){\makebox(0,0)[lb]{\raisebox{0pt}[0pt][0pt]{\shortstack[l]{{\twlrm
$m_2$}}}}}
\put(150,0){\makebox(0,0)[lb]{\raisebox{0pt}[0pt][0pt]{\shortstack[l]{{\twlrm
$l_2=m_1=p_1$}}}}}
\put(230,95){\makebox(0,0)[lb]{\raisebox{0pt}[0pt][0pt]{\shortstack[l]{{\twlrm
$n_3$}}}}}
\put(230,55){\makebox(0,0)[lb]{\raisebox{0pt}[0pt][0pt]{\shortstack[l]{{\twlrm
$n_2$}}}}}
\end{picture}
\noindent
The figure shows the composition of $A^{(i+1)}_4(\zeta)$ and
$\Phi^{(i+1,i)}(\zeta)$.  The $({\bf p},{\bf m})$-component of
$A^{(i+1)}_4(\zeta)$ is represented by the part A and the
$({\bf m},{\bf n})$-component of $\Phi^{(i+1,i)}(\zeta)$ is represented by the
part B.  The composition of A followed by B
(with $(m_2,m_3,\ldots)$ summed over)
can be viewed as the $({\bf l},{\bf n})$-component of $A^{(i)}_4(\zeta)$,
except that the
product $A^{(i+1)}_4(\zeta)\Phi^{(i+1,i)}(\zeta)$ is acting as
$\H^{(i)}\rightarrow\H^{(i+1)}$ while $A^{(i)}_4(\zeta)$ as
$\H^{(i)}\rightarrow\H^{(i)}$.  To obtain an equality, $\H^{(i)}$ and
$\H^{(i+1)}$ must be intertwined.
In view of the initial condition (\ref{eqn:init}) for the Boltzmann weights,
the required intertwining action
is exactly what $\Phi^{(i+1,i)}(1)$ does. Therefore, we have
\begin{eq}
A^{(i+1)}_4(\zeta)_{l_2}\Phi^{(i+1,i)}(\zeta)_{l_2,l_1}&=&
\Phi^{(i+1,i)}(1)_{l_2,l_1}A^{(i)}_4(\zeta)_{l_1}.\nonumber
\end{eq}

The normalization (\ref{eqn:Facedual}) can be shown using
unitarity (\ref{eqn:uni}) and crossing symmetry (\ref{eqn:cross}).
To motivate (\ref{eqn:Facecomm}) we make use of the Yang-Baxter equation.

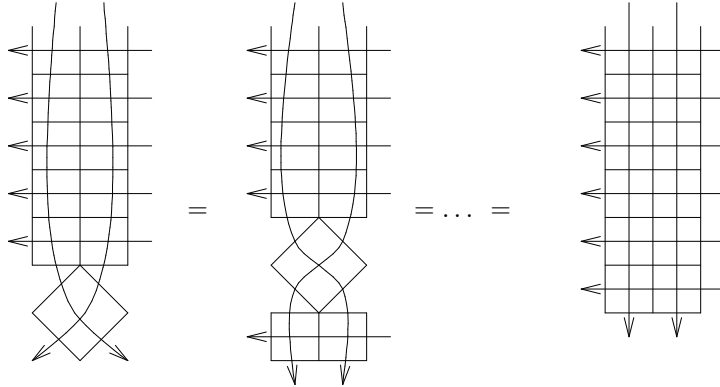
\setlength{\unitlength}{0.0125in}
\begin{picture}(300,175)(0,-10)
\drawline(300,140)(240,140)
\drawline(248.000,142.000)(240.000,140.000)(248.000,138.000)
\drawline(300,120)(240,120)
\drawline(248.000,122.000)(240.000,120.000)(248.000,118.000)
\drawline(300,100)(240,100)
\drawline(248.000,102.000)(240.000,100.000)(248.000,98.000)
\drawline(300,80)(240,80)
\drawline(248.000,82.000)(240.000,80.000)(248.000,78.000)
\drawline(300,60)(240,60)
\drawline(248.000,62.000)(240.000,60.000)(248.000,58.000)
\drawline(300,40)(240,40)
\drawline(248.000,42.000)(240.000,40.000)(248.000,38.000)
\drawline(280,160)(280,20)
\drawline(278.000,28.000)(280.000,20.000)(282.000,28.000)
\drawline(260,160)(260,20)
\drawline(258.000,28.000)(260.000,20.000)(262.000,28.000)
\drawline(250,130)(290,130)
\drawline(250,110)(290,110)
\drawline(250,90)(290,90)
\drawline(250,70)(290,70)
\drawline(250,50)(290,50)
\drawline(270,150)(270,30)
\drawline(250,150)(250,30)(290,30)(290,150)
\drawline(160,140)(100,140)
\drawline(108.000,142.000)(100.000,140.000)(108.000,138.000)
\drawline(160,120)(100,120)
\drawline(108.000,122.000)(100.000,120.000)(108.000,118.000)
\drawline(160,100)(100,100)
\drawline(108.000,102.000)(100.000,100.000)(108.000,98.000)
\drawline(160,80)(100,80)
\drawline(108.000,82.000)(100.000,80.000)(108.000,78.000)
\drawline(160,20)(100,20)
\drawline(108.000,22.000)(100.000,20.000)(108.000,18.000)
\drawline(110,90)(150,90)
\drawline(110,110)(150,110)
\drawline(110,130)(150,130)
\drawline(130,150)(130,70)
\drawline(110,150)(110,70)(150,70)(150,150)
\drawline(130,70)(110,50)(130,30)
	(150,50)(130,70)
\drawline(130,30)(130,10)
\drawline(110,30)(110,10)(150,10)
	(150,30)(110,30)
\drawline(60,140)(0,140)
\drawline(8.000,142.000)(0.000,140.000)(8.000,138.000)
\drawline(60,120)(0,120)
\drawline(8.000,122.000)(0.000,120.000)(8.000,118.000)
\drawline(60,100)(0,100)
\drawline(8.000,102.000)(0.000,100.000)(8.000,98.000)
\drawline(60,80)(0,80)
\drawline(8.000,82.000)(0.000,80.000)(8.000,78.000)
\drawline(60,60)(0,60)
\drawline(8.000,62.000)(0.000,60.000)(8.000,58.000)
\drawline(10,70)(50,70)
\drawline(10,90)(50,90)
\drawline(10,110)(50,110)
\drawline(10,130)(50,130)
\drawline(30,150)(30,50)
\drawline(10,150)(10,50)(50,50)(50,150)
\drawline(30,50)(10,30)(30,10)
	(50,30)(30,50)
\drawline(140,160)	(140.357,157.383)
	(140.703,154.812)
	(141.037,152.285)
	(141.361,149.802)
	(141.973,144.967)
	(142.540,140.301)
	(143.061,135.801)
	(143.537,131.461)
	(143.968,127.278)
	(144.354,123.245)
	(144.694,119.360)
	(144.989,115.617)
	(145.238,112.011)
	(145.442,108.539)
	(145.601,105.195)
	(145.714,101.975)
	(145.782,98.875)
	(145.805,95.890)
	(145.782,93.014)
	(145.714,90.244)
	(145.601,87.576)
	(145.442,85.004)
	(144.989,80.131)
	(144.354,75.589)
	(143.537,71.342)
	(142.540,67.352)
	(141.361,63.584)
	(140.000,60.000)

\drawline(140,60)	(138.422,57.406)
	(136.054,54.893)
	(133.159,52.433)
	(130.000,50.000)
	(126.841,47.567)
	(123.946,45.107)
	(121.578,42.594)
	(120.000,40.000)

\drawline(120,40)	(118.984,37.059)
	(118.258,33.765)
	(117.823,29.999)
	(117.678,25.645)
	(117.714,23.210)
	(117.823,20.585)
	(118.005,17.754)
	(118.258,14.702)
	(118.585,11.416)
	(118.984,7.880)
	(119.456,4.079)
	(120.000,0.000)

\drawline(116.929,7.653)(120.000,0.000)(120.892,8.198)
\drawline(120,160)	(119.643,157.383)
	(119.297,154.812)
	(118.963,152.285)
	(118.639,149.802)
	(118.027,144.967)
	(117.460,140.301)
	(116.939,135.801)
	(116.463,131.461)
	(116.032,127.278)
	(115.646,123.245)
	(115.306,119.360)
	(115.011,115.617)
	(114.762,112.011)
	(114.558,108.539)
	(114.399,105.195)
	(114.286,101.975)
	(114.218,98.875)
	(114.195,95.890)
	(114.218,93.014)
	(114.286,90.244)
	(114.399,87.576)
	(114.558,85.004)
	(115.011,80.131)
	(115.646,75.589)
	(116.463,71.342)
	(117.460,67.352)
	(118.639,63.584)
	(120.000,60.000)

\drawline(120,60)	(121.578,57.406)
	(123.946,54.893)
	(126.841,52.433)
	(130.000,50.000)
	(133.159,47.567)
	(136.054,45.107)
	(138.422,42.594)
	(140.000,40.000)

\drawline(140,40)	(141.016,37.059)
	(141.742,33.765)
	(142.177,29.999)
	(142.322,25.645)
	(142.286,23.210)
	(142.177,20.585)
	(141.995,17.754)
	(141.742,14.702)
	(141.415,11.416)
	(141.016,7.880)
	(140.544,4.079)
	(140.000,0.000)

\drawline(139.108,8.198)(140.000,0.000)(143.071,7.653)
\drawline(40,160)	(40.287,156.704)
	(40.563,153.466)
	(40.829,150.284)
	(41.083,147.159)
	(41.326,144.089)
	(41.557,141.073)
	(41.778,138.111)
	(41.988,135.202)
	(42.186,132.345)
	(42.373,129.541)
	(42.549,126.787)
	(42.714,124.083)
	(42.867,121.429)
	(43.009,118.824)
	(43.139,116.267)
	(43.258,113.757)
	(43.462,108.877)
	(43.620,104.178)
	(43.732,99.654)
	(43.797,95.299)
	(43.816,91.109)
	(43.787,87.076)
	(43.712,83.196)
	(43.589,79.463)
	(43.419,75.870)
	(43.201,72.412)
	(42.935,69.085)
	(42.620,65.880)
	(42.257,62.794)
	(41.846,59.820)
	(41.385,56.953)
	(40.876,54.187)
	(40.316,51.515)
	(39.708,48.934)
	(38.340,44.015)
	(36.772,39.386)
	(35.000,35.000)

\drawline(35,35)	(32.142,30.102)
	(30.048,27.498)
	(27.413,24.693)
	(24.164,21.614)
	(20.228,18.188)
	(15.531,14.341)
	(12.874,12.237)
	(10.000,10.000)

\drawline(15.109,16.473)(10.000,10.000)(17.554,13.307)
\drawline(20,160)	(19.713,156.704)
	(19.437,153.466)
	(19.171,150.284)
	(18.917,147.159)
	(18.674,144.089)
	(18.443,141.073)
	(18.222,138.111)
	(18.012,135.202)
	(17.814,132.345)
	(17.627,129.541)
	(17.451,126.787)
	(17.286,124.083)
	(17.133,121.429)
	(16.991,118.824)
	(16.861,116.267)
	(16.742,113.757)
	(16.538,108.877)
	(16.380,104.178)
	(16.268,99.654)
	(16.203,95.299)
	(16.184,91.109)
	(16.213,87.076)
	(16.288,83.196)
	(16.411,79.463)
	(16.581,75.870)
	(16.799,72.412)
	(17.065,69.085)
	(17.380,65.880)
	(17.743,62.794)
	(18.154,59.820)
	(18.615,56.953)
	(19.124,54.187)
	(19.684,51.515)
	(20.292,48.934)
	(21.660,44.015)
	(23.228,39.386)
	(25.000,35.000)

\drawline(25,35)	(27.858,30.102)
	(29.952,27.498)
	(32.587,24.693)
	(35.836,21.614)
	(39.772,18.188)
	(44.469,14.341)
	(47.126,12.237)
	(50.000,10.000)

\drawline(42.446,13.307)(50.000,10.000)(44.891,16.473)
\put(170,70){\makebox(0,0)[lb]{\raisebox{0pt}[0pt][0pt]{\shortstack[l]{{\twlrm
$=\ldots\ =$}}}}}
\put(75,70){\makebox(0,0)[lb]{\raisebox{0pt}[0pt][0pt]{\shortstack[l]{{\twlrm
$=$}}}}}
\end{picture}
\noindent
In the above,
the face added in the first figure is pushed up to infinity,
using the YB equations, and
because of the normalization by $\kappa$ its {\em finite} effect will
disappear in the {\em infinite} limit.

Now let us relate the VO's to the correlation functions.
As an illustration we discuss the simplest situation.
Look at the figure below. We wish to calculate the
local probability of the adjacent variables $\l_1$ and $\l_2$
taking the values $l_1$ and $l_2$, respectively.
We denote this probability by $P^{(i)}(l_1,l_2)$.

Divide the whole lattice into 6 pieces. The main parts are the CTM's.
The remaining pieces give the graphical definition of two types of VO's
\[
\Phi^{(i+1,i)}_1(\zeta):\H^{(i)}\goto{}\H^{(i+1)}
\quad
\Phi^{(i,i+1)}_2(\zeta):\H^{(i+1)}\goto{}\H^{(i)}.
\]

\setlength{\unitlength}{0.0125in}
\begin{picture}(320,325)(0,-10)
\drawline(140,65)(180,65)
\drawline(140,105)(180,105)
\drawline(140,205)(180,205)
\drawline(140,245)(180,245)
\drawline(140,145)(180,145)(180,25)
	(140,25)(140,145)
\drawline(140,285)(180,285)(180,165)
	(140,165)(140,285)
\drawline(120,25)(80,25)(80,145)
	(120,145)(120,25)
\drawline(120,65)(40,65)(40,145)
\drawline(120,105)(0,105)
\drawline(0,105)(0,145)(80,145)
\drawline(320,105)(320,145)(240,145)
\drawline(200,105)(320,105)
\drawline(200,65)(280,65)(280,145)
\drawline(200,25)(240,25)(240,145)
	(200,145)(200,25)
\drawline(0,205)(0,165)(80,165)
\drawline(120,205)(0,205)
\drawline(120,245)(40,245)(40,165)
\drawline(120,285)(80,285)(80,165)
	(120,165)(120,285)
\drawline(200,285)(240,285)(240,165)
	(200,165)(200,285)
\drawline(200,245)(280,245)(280,165)
\drawline(200,205)(320,205)
\drawline(320,205)(320,165)(240,165)
\put(10,265){\makebox(0,0)[lb]{\raisebox{0pt}[0pt][0pt]{\shortstack[l]{{\twlrm
$A^{(i+1)}_4(\zeta)$}}}}}
\put(130,150){\makebox(0,0)[lb]{\raisebox{0pt}[0pt][0pt]{\shortstack[l]{{\twlrm
$\l_2$}}}}}
\put(185,150){\makebox(0,0)[lb]{\raisebox{0pt}[0pt][0pt]{\shortstack[l]{{\twlrm
$\l_1$}}}}}
\put(150,0){\makebox(0,0)[lb]{\raisebox{0pt}[0pt][0pt]{\shortstack[l]{{\twlrm
$\Phi^{(i,i+1)}_2(\zeta)$}}}}}
\put(150,300){\makebox(0,0)[lb]{\raisebox{0pt}[0pt][0pt]{\shortstack[l]{{\twlrm
$\Phi^{(i+1,i)}_1(\zeta)$}}}}}
\put(290,265){\makebox(0,0)[lb]{\raisebox{0pt}[0pt][0pt]{\shortstack[l]{{\twlrm
$A^{(i)}_1(\zeta)$}}}}}
\put(290,45){\makebox(0,0)[lb]{\raisebox{0pt}[0pt][0pt]{\shortstack[l]{{\twlrm
$A^{(i)}_2(\zeta)$}}}}}
\put(10,45){\makebox(0,0)[lb]{\raisebox{0pt}[0pt][0pt]{\shortstack[l]{{\twlrm
$A^{(i+1)}_3(\zeta)$}}}}}
\end{picture}
\noindent

The operator
$\Phi^{(i+1,i)}_1(\zeta)$ is the same as
$\Phi^{(i+1,i)}(\zeta)$ that we introduced.
In much the same way as for the CTM
(\ref{eqn:CTM3})-(\ref{eqn:CTM1}), $\Phi^{(i,i+1)}_2(\zeta)$ can be written
in terms of $\Phi^{(i+1,i)}(\zeta)$.
Namely using the crossing symmetry (\ref{eqn:cross}) we have
\begin{eq}
\Phi^{(i,i+1)}_2(\zeta)&\sim&R^{(i)}
\Phi^{(i,i+1)}(\zeta)G^{(i+1)}_1R^{(i+1)-1},\label{eqn:VOcross}
\end{eq}
where $G^{(i)}_1$ and $R^{(i)}$ are defined in (\ref{eqn:opSG}) and
(\ref{eqn:opR}).

Using these operators, we have
\begin{eq}
P^{(i)}(l_1,l_2)&=& {Q^{(i)}(l_1,l_2)
\over \sum_{l'_1,l'_2}Q^{(i)}(l'_1,l'_2)},\label{eqn:probab}\\
Q^{(i)}(l_1,l_2)&=&\tr_{\H^{(i)}_{l_1}}\Bigl(
A^{(i)}_1(\zeta)_{l_1}A^{(i)}_2(\zeta)_{l_1}\times\nonumber\\
&&\hskip-1cm\Phi^{(i,i+1)}_2(\zeta)_{l_1,l_2}
A^{(i+1)}_3(\zeta)_{l_2}A^{(i+1)}_4(\zeta)_{l_2}
\Phi^{(i+1,i)}_1(\zeta)_{l_2,l_1}\Bigr). \label{eqn:Q}
\end{eq}
By (\ref{eqn:Facehomo}) and (\ref{eqn:VOcross})
we can change the order of the product
$$\Phi^{(i,i+1)}_2(\zeta)_{l_1,l_2}A^{(i+1)}_3(\zeta)_{l_2}
A^{(i+1)}_4(\zeta)_{l_2}$$ in (\ref{eqn:Q}),
to find
\begin{eq}
Q^{(i)}(l_1,l_2)&=&g_{l_1}g_{l_2}\times\nonumber\\
&&\hskip-2cm \tr_{\H^{(i)}_{l_1}} \Bigl(x^{2D^{(i)}_{l_1}}
\Phi^{(i,i+1)}(x\zeta)_{l_1,l_2}\Phi^{(i+1,i)}(\zeta)_{l_2,l_1}\Bigr).
\end{eq}
If we choose the normalization (\ref{eqn:Facedual})
then the denominator $\sum_{l_1,l_2}Q^{(i)}(l_1,l_2)$ in (\ref{eqn:probab})
 is equal to
$\tr_{\H^{(i)}}\Bigl(G^{(i)}_1x^{2D^{(i)}}\Bigr)$.
In a similar way, it is straightforward to relate the general correlators
to traces of products of VO's.

\subsection{The $q$-difference equation}

Suppose that $n$ is even and the one-dimensional configuration
$[l_1,l_2,\ldots,l_n,l_{n+1}]$
is admissible: $l_{k+1}=l_k\pm1$, and
$l_{n+1}=l_1$. We call such a configuration a cyclic path of length $n$
and denote it by $[l_1,\ldots,l_n]$.
Consider the vector space ${\cal V}_n$ having the cyclic
paths of length $n$ as basis.
If $l_1,\ldots,l_n$ do not satisfy
the above admissibility
condition we regard $[l_1,\ldots,l_n]$ to be a null vector.
On ${\cal V}_n$ we define an operator $W_k(\zeta)\ (2\le k\le n)$ by
\begin{eq}
W_k(\zeta)[l_1,\ldots,l_n]
&=&\sum_{l'_k}\w(l'_k,l_{k+1},l_k,l_{k-1},\zeta)
[l_1,\ldots,l_{k-1},l'_k,l_{k+1},\ldots,l_n].\nonumber\\
\end{eq}

Define further
\begin{eq}
F^{(i)}_{l_1,\ldots,l_n}(\zeta_1,\ldots,\zeta_n)
&=&\tr_{\H^{(i)}_{l_1}}\Bigl(x^{2D^{(i)}_{l_1}}
\Phi^{(i,i+1)}(\zeta_1)_{l_1,l_2}\cdots
\Phi^{(i+1,i)}(\zeta_n)_{l_n,l_1}\Bigr).\nonumber\\
\end{eq}
The commutation relation (\ref{eqn:Facecomm}) implies
\begin{eq}
&&\sum_{l'_{k+1}}\w(l_{k+1},l_{k+2},l'_{k+1},l_k,\zeta_k/\zeta_{k+1})
F^{(i)}_{l_1,\ldots,l_k,l'_{k+1},l_{k+2},\ldots,l_n}
(\zeta_1,\ldots,\zeta_k,\zeta_{k+1},\ldots,\zeta_n)\nonumber\\
&&=F^{(i)}_{l_1,\ldots,l_n}
(\zeta_1,\ldots,\zeta_{k+1},\zeta_k,\ldots,\zeta_n).
\label{eqn:symm}
\end{eq}
If we set
\begin{eq}
F^{(i)}(\zeta_1,\ldots,\zeta_n)&=&
\sum_{l_1,\ldots,l_n}F^{(i)}_{l_1,\ldots,l_n}(\zeta_1,\ldots,\zeta_n)
[l_1,\ldots,l_n],
\end{eq}
then (\ref{eqn:symm}) reads as
\begin{eq}
W_{k+1}(\zeta_k/\zeta_{k+1})F^{(i)}(\zeta_1,\ldots,\zeta_n)
&=&
F^{(i)}(\zeta_1,\ldots,\zeta_{k+1},\zeta_k,\ldots,\zeta_n).
\end{eq}

Using the cyclic property of the
trace and the commutation relations we have the difference equation:

\begin{eq}
F^{(i)}(\zeta_1,\ldots,x^2\zeta_k,\ldots,\zeta_n)
&=&W_k(\zeta_{k-1}/x^2\zeta_k)^{-1}\cdots W_2(\zeta_1/x^2\zeta_k)^{-1}\times
\nonumber\\
&&\hskip-2cmCW_n(\zeta_k/\zeta_n)\cdots W_{k+1}(\zeta_k/\zeta_{k+1})
F^{(i+1)}(\zeta_1,\ldots,\zeta_n),
\end{eq}
where $C$ is such that $C[l_1,\ldots,l_n]=[l_n,l_1,\ldots,l_{n-1}]$.

\subsection{Construction by representation theory}

We now proceed to the mathematical construction of VO's.
We shall use the following notations \cite{JMOh}:

$U=U_{q}\bigl(\widehat{\goth{sl}_2}\bigr)$ denotes the quantized affine
algebra of type $A^{(1)}_1$ with $q=-x$ as the deformation parameter.

The Chevalley generators are written as
$e_i, f_i, t_i=q^{h_i}$ ($i=0,1$) and $q^d$.

$U'$ will denote the subalgebra of $U$ generated by $e_i, f_i, t_i$ ($i=0,1$).

$V(\lambda)$ is an irreducible $U$-module with highest
weight $\lambda$ and highest weight vector $\ket{\lambda}$.
$\Lambda_0, \Lambda_1$ denote the fundamental weights.

$V$ is a two-dimensional representation of
$U_q\bigl(\goth{sl}_2\bigr)$ with a standard basis: $\{v_{+},\ v_{-}\}$, and
$V_z$ denotes the affinization of $V$ with the spectral parameter $z$.
Note that the grading operator $d$ has a well-defined action on $V(\lambda)$ or
$V_z$.
Finally $P\Rb(z)$ is the $R$-matrix, i.e. the intertwiner
$P\Rb(z_1/z_2):V_{z_1}\otimes V_{z_2}\rightarrow V_{z_2}\otimes V_{z_1}$,
which is normalized in such a way that it sends $v_+\otimes v_+$ to itself.

In contrast with the situation in the case of the XXZ spin chain, and
the 8-vertex models, here we are interested in VO's
of the face-type. The building blocks of the latter are
still the VO's of the vertex type:

\begin{eq}
\Phit^{\mu\,V}_\lambda(z)&:& V(\lambda) \longrightarrow V(\mu)\otimes V_z.
\label{eqn:typeI}
\end{eq}
Here $\lambda,\mu$ are dominant integral weights of level $k\ge 1$.
\thanks{Strictly speaking we need to complete the spaces properly;
but in the sequel we shall not dwell upon this point
(see \cite{DJO}, \cite{DFJMN}).}
A non-trivial VO of vertex type (\ref{eqn:typeI}) exists only
when $\mu=\lambda_{\pm}\bydef\lambda\pm (\Lambda_1-\Lambda_0)$,
in which case it is unique up to a scalar.
We normalize them as
\[
\Phit^{\lambda_\pm}_\lambda(z)_{\mp}\ket{\lambda}
=\ket{\lambda_{\pm}}+O(z)
\]
where the components of (\ref{eqn:typeI}) are defined by
$\Phit^{\mu\,V}_\lambda(z)
=\sum_\vep \Phit^{\mu}_\lambda(z)_{\vep}\otimes v_{\vep}$.
They enjoy the basic properties (\cite{JMOh}, Appendix 2):
\begin{eq}
&&z^d\circ \Phit^{\mu}_\lambda(z')_\vep\circ z^{-d}
=\Phit^{\mu}_\lambda(z'/z)_\vep,  \label{eqn:typeIhomo} \\
&&P\Rb(z_1/z_2)\Phit^{\nu\, V}_{\mu}(z_1)\Phit^{\mu\,V}_{\lambda}(z_2) \n
&&=
\sum_{\mu'}\Phit^{\nu\, V}_{\mu'}(z_2)\Phit^{\mu'\,V}_{\lambda}(z_1)
\overline{W}^1_k
\left(\matrix{\lambda&\mu\cr \mu'&\nu\cr}\Bigm|\,z_1/z_2\right)
{\psi_k(p z_2/z_1) \over \psi_k(p z_1/z_2)},
\label{eqn:typeIcomm}\\
&&\sum_\vep x^{(\pm1+\vep)/2} \,\Phit^{\nu}_{\lambda_\pm}(x^{-2}z)_{-\vep}
\Phit^{\lambda_\pm}_\lambda(z)_\vep
=\delta_{\lambda\nu}g^{\lambda_\pm}_\lambda
\times \id_{V(\lambda)}.\label{eqn:typeIdual}
\end{eq}
Here we have set $p=x^{2(k+2)}$,
\begin{eq}
\psi_k(z)&=&{(x^4z;x^4,p)_\infty (z;x^4,p)_\infty
\over (x^2z;x^4,p)_\infty^2}, \n
g^{\lambda_{\pm}}_{\lambda}&=&
{1 \over \psi_k(p)}
{(p^{r_\mp}x^2;p)_\infty \over(p^{r_\mp};p)_\infty},
\end{eq}
and if $\lambda=(k-a)\Lambda_0+a\Lambda_1$
with $\Lambda_0, \Lambda_1$ being the fundamental weights,
then $r_-=(a+1)/(k+2)$, $r_+=1-r_-$.
The formulas for $\overline{W}^1_k$ are given in \cite{JMOh}, eq.(A2.2)
wherein $q=-x$.

In the particular case where $\lambda$ has level $k=1$,
(\ref{eqn:typeI}) gives rise to an isomorphism \cite{DJO},\cite{DFJMN}.
In this case (\ref{eqn:typeIcomm}) specializes to
\begin{eq}
&&P\Rb(z_1/z_2)\Phit^{\Lambda_i\, V}_{\Lambda_{i+1}}(z_1)
\Phit^{\Lambda_{i+1}\,V}_{\Lambda_i}(z_2) \n
&&=\Phit^{\Lambda_i\, V}_{\Lambda_{i+1}}(z_2)
\Phit^{\Lambda_{i+1}\,V}_{\Lambda_i}(z_1)
z^{1-i}{(x^{4}z;x^4)_\infty \over (x^{4}z^{-1};x^4)_\infty}
{(x^{2}z^{-1};x^4)_\infty \over (x^{2}z;x^4)_\infty}
\quad (z=z_1/z_2),  \nonumber
\end{eq}
and $g^{\Lambda_{i+1}}_{\Lambda_i}=g^{\Lambda_{i}}_{\Lambda_{i+1}}
={(x^4;x^4)_\infty/(x^2;x^4)_\infty}$.
Here and till the end of this subsection, $i$ is understood to be either
$0$ or $1$.

\medskip
\noindent{\sl Erratum.}\quad
In \cite{JMOh}, the $r_\pm$ in the right hand sides of eqs.(A2.3--4)
are to be corrected to $r_\mp$.
\bigskip

Fix $L = k + 2$,
$\xi=(L-m-2)\Lambda_0+(m-1)\Lambda_1$.
Set $\lambda_l=(L -l -1)\Lambda_0+(l-1)\Lambda_1$.

Define
\begin{eq}
\bH^{(i)}&=&\oplus_{l=1}^{k+1}\bH^{(i)}_l,\n
\bH^{(i)}_l&=&\hom\left(V(\lambda_l),V(\xi)\otimes V(\Lambda_{i})\right).
\label{eqn:H}
\end{eq}
Namely, $\bar{\cal H}^{(i)}_l$ is the space of all
{\it $U'$-linear} maps (i.e. those commuting with the action of $U'$)
$f:V(\lambda_l)\rightarrow V(\xi)\otimes V(\Lambda_i)$.
Because of the $U'$-linearity, $f$ is completely determined by the image, say
$v=f(u_{\lambda_l})$, of the highest weight vector $u_{\lambda_l}$ in
$V(\lambda_l)$. Conversely, any vector $v$ in $V(\xi)\otimes V(\Lambda_i)$,
such that it is killed by the $e_i$ and has the weight $\lambda_l$
(modulo the null root), gives rise to such a $U'$-linear map.
Thus we have the identification
\[
\bH^{(i)}_{l}=
\{ v\in V(\xi)\otimes V(\Lambda_{i}) \mid
e_j v=0,~t_j v=q^{\br{h_j,\lambda_l}}v\quad \forall j=0,1\}.
\]
The definition of the vertex
operators of face type is best described in the language of maps, rather
than vectors.

The VO's of face type,
\[
\Phit^{(i+1,i)}(z)_{l',l}
{}~:~ \bH^{(i)}_l \longrightarrow \bH^{(i+1)}_{l'},
\]
are defined as follows.
Given $f\in \bH^{(i)}_l$, consider the composition
\begin{eq}
V(\lambda_l)&\goto{f}  &V(\xi)\otimes V(\Lambda_i) \n
&\goto{\scriptstyle id \otimes \Phit^{\Lambda_{i+1}V}_{\Lambda_i}(z)}
& V(\xi)\otimes V(\Lambda_{i+1})\otimes V_z   \n
&\simeq&\left(\oplus_{l'=1}^{L-1} \bH^{(i+1)}_{l'}
\otimes V(\lambda_{l'})\right)
\otimes V_z.
\label{eqn:dec}
\end{eq}
{}From (\ref{eqn:dec})
it follows that there exist unique maps
$f_{l'}\in \bH^{(i+1)}_{l'}$ such that the following is true:
\begin{eqnarray}
&&\left(\id_{V(\xi)}\otimes \Phit^{\Lambda_{i+1}\,V}_{\Lambda_i}(z)\right)
\circ f
=\sum_{1\le l'\le L-1} \left(f_{l'}\otimes \id_V\right)
\circ \Phit^{\lambda_{l'}\,V}_{\lambda_l}(z).\label{eqn:VOface1}
\end{eqnarray}
We define $\Phit^{(i+1,i)}(z)_{l',l}$ by
\begin{eq}
&&\Phit^{(i+1,i)}(z)_{l',l}(f)=f_{l'}. \label{eqn:VOface2}
\end{eq}
It is immediate to see that $\Phit^{(i+1,i)}(z)_{l',l}$
is nonzero only when $l'=l\pm 1$ and $1\le l' \le L-1$.

The following relations are simple consequences of
the defining relations (\ref{eqn:VOface1}), (\ref{eqn:VOface2})
and the corresponding properties
(\ref{eqn:typeIhomo}),
(\ref{eqn:typeIcomm}),
(\ref{eqn:typeIdual}) of the VO of vertex type.
\begin{eq}
&&z^d\circ \Phit^{(i+1,i)}(z')_{l',l}\circ z^{-d}
=\Phit^{(i+1,i)}(z'/z)_{l',l}, \label{eqn:facehomo} \\
&&\Phit^{(i,i+1)}(z_2)_{l_4,l_1}\Phit^{(i+1,i)}(z_1)_{l_1,l_2}
=\sum_{l_3}
\Phit^{(i,i+1)}(z_1)_{l_4,l_3}
\Phit^{(i+1,i)}(z_2)_{l_3,l_2} \n
&&\quad\times \left({z_1 \over z_2}\right)^{i-1}
\wt(l_1,l_2,l_3,l_4,z_1/z_2), \label{eqn:facecomm} \\
&&\sum_{\pm} x^{(1\mp 1)/2}g^{\lambda_{l\pm1}}_{\lambda_l}
\Phit^{(i,i+1)}(zx^{-2})_{l,l\pm 1}
\Phit^{(i+1,i)}(z)_{l\pm 1,l}
=x^{i}g^{\Lambda_{i+1}}_{\Lambda_i}\times \id.
\end{eq}
Here
\begin{eq}
\wt(l_1,l_2,l_3,l_4,z)
&=&{\wb(\lambda_{l_1},\lambda_{l_4},\lambda_{l_3},\lambda_{l_2},z,k)
\over
\wb({\Lambda_{i+1}},{\Lambda_i},{\Lambda_{i+1}},{\Lambda_i},z,1)
}
{\psi_k(pz^{-1}) \over \psi_k(pz)}{\psi_1(x^6z) \over \psi_1(x^6z^{-1})}
z^{1-i} \n
&=&{1\over {\bar \kappa}(z^{-1/2})}
\wb(\lambda_{l_1},\lambda_{l_4},\lambda_{l_3},\lambda_{l_2},z,k)
\end{eq}
where $k=L-2$, $p=x^{2L}$, and
${\bar \kappa}(\zeta)$ is given in (\ref{eqn:kappa}).
{}From the formulas (A2.2), \cite{JMOh} one can verify that
\begin{eq}
&&\wt(l_1,l_2,l_3,l_4,\zeta^{-2})
=\w(l_1,l_2,l_3,l_4,\zeta)
{F_{l_4l_3}F_{l_3l_2} \over F_{l_4l_1}F_{l_1l_2}}
\zeta^{(l_4+l_2-l_3-l_1)/2-1},\n
&&F_{l\,l+1}=\Gamma_p({l\over L}),
\quad F_{l\,l-1}=\Gamma_p(1-{l\over L}).
\end{eq}
Here we used (\ref{eqn:bol1})-(\ref{eqn:bol3}).

Now, we return to the ABF models.
Let $h \ (h\in\C h_0\oplus\C h_1\oplus\C d)$ act on $\bH^{(i)}$ by
the adjoint action, $f\in\bH^{(i)}\rightarrow h\circ f-f\circ h$.
It is known that the principally specialized character of $\bH^{(i)}_l$,
i.e., $\tr_{\bH^{(i)}_l}x^{-2\rho}$ where $\rho=\Lambda_0+\Lambda_1$,
is equal to $\tr_{\H^{(i)}_l}x^{2D^{(i)}}$. Therefore, we identify
$\H^{(i)}_l$ with $\bH^{(i)}_l$ and $D^{(i)}$ with $-\rho$. Furthermore,
with the identification
\[
\Phi^{(ii')}(\zeta)_{ll'}
=
c^{-1}F_{ll'} \zeta^{(i'-i+l-l')/2}\,\Phit^{(i\,i')}(\zeta^{-2})_{ll'}
\]
we find that the properties (\ref{eqn:Facehomo}),
(\ref{eqn:Facecomm}),
(\ref{eqn:Facedual})
are satisfied by $\Phi^{(ii')}(\zeta)_{ll'}$
where
$c^2=(p;x^4,p)_\infty^2(x^4;x^4)_\infty(p;p)_\infty^2/
(px^2;x^4,p)_\infty^2(x^2;x^4)_\infty\times (1-p)^{(L+1)/L}$.

\subsection{The Kashiwara-Miwa model}

In this subsection, we wish to give a brief treatment
of the Kashiwara-Miwa (KM) series as an example of
edge-interaction models \cite{KM,JMO}. The discussion
of the VO's in the previous subsections can immediately be applied
here, since an edge-interaction model can be regarded
as a special case of a face-interaction model.

Take an integer $N\ge2$. If $N$ is even (resp. odd),
set $N=2n$ (resp. $2n+1$). Consider the same square lattice as
in the ABF model. For the KM model, we consider a local variable
$\l_j$ taking its values in $\Z/N\Z\sqcup\{\bullet\}$.
For $a\in\Z/N\Z$ we define an integer $a^*$ such that
$0\le a^*\le n$, $a^*\equiv a\hbox{ or } -a$.
The admissibility condition we impose on
$(\l_j,\l_{j'})$ is such that one and only one of
$\l_j,\l_{j'}$ takes the value $\bullet$.
Thus a half of the variables are `frozen' to the state $\bullet$.
The Boltzmann weight
is given as follows:

\setlength{\unitlength}{0.0125in}
\begin{picture}(148,230)(0,-10)
\drawline(105,55)(25,55)
\drawline(33.000,57.000)(25.000,55.000)(33.000,53.000)
\drawline(65,95)(65,15)
\drawline(63.000,23.000)(65.000,15.000)(67.000,23.000)
\drawline(45,75)(85,75)(85,35)
	(45,35)(45,75)
\drawline(45,195)(85,195)(85,155)
	(45,155)(45,195)
\drawline(65,215)(65,135)
\drawline(63.000,143.000)(65.000,135.000)(67.000,143.000)
\drawline(105,175)(25,175)
\drawline(33.000,177.000)(25.000,175.000)(33.000,173.000)
\put(130,55){\makebox(0,0)[lb]{\raisebox{0pt}[0pt][0pt]{\shortstack[l]{{\twlrm
$\w(a,\bullet,b,\bullet,\zeta)
=\w(\bullet,b,\bullet,a,x/\zeta)
{\displaystyle\frac{g_b}{g_\bullet}}$}}}}}
\put(130,170){\makebox(0,0)[lb]{\raisebox{0pt}[0pt][0pt]{\shortstack[l]{{\twlrm
$\w(\bullet,b,\bullet,a,\zeta)
={\displaystyle\frac{\zeta^{|a^{*}-b^{*}|}}{\kappa(\zeta)}h(\zeta;a,b)}$}}}}}
\put(95,80){\makebox(0,0)[lb]{\raisebox{0pt}[0pt][0pt]{\shortstack[l]{{\twlrm
$b$}}}}}
\put(30,20){\makebox(0,0)[lb]{\raisebox{0pt}[0pt][0pt]{\shortstack[l]{{\twlrm
$a$}}}}}
\put(95,145){\makebox(0,0)[lb]{\raisebox{0pt}[0pt][0pt]{\shortstack[l]{{\twlrm
$b$}}}}}
\put(35,200){\makebox(0,0)[lb]{\raisebox{0pt}[0pt][0pt]{\shortstack[l]{{\twlrm
$a$}}}}}
\put(90,20){\makebox(0,0)[lb]{\raisebox{0pt}[0pt][0pt]{\shortstack[l]{{\twlrm
$\bullet$}}}}}
\put(40,80){\makebox(0,0)[lb]{\raisebox{0pt}[0pt][0pt]{\shortstack[l]{{\twlrm
$\bullet$}}}}}
\put(40,145){\makebox(0,0)[lb]{\raisebox{0pt}[0pt][0pt]{\shortstack[l]{{\twlrm
$\bullet$}}}}}
\put(85,200){\makebox(0,0)[lb]{\raisebox{0pt}[0pt][0pt]{\shortstack[l]{{\twlrm
$\bullet$}}}}}
\put(0,55){\makebox(0,0)[lb]{\raisebox{0pt}[0pt][0pt]{\shortstack[l]{{\twlrm
$\zeta_H$}}}}}
\put(60,0){\makebox(0,0)[lb]{\raisebox{0pt}[0pt][0pt]{\shortstack[l]{{\twlrm
$\zeta_V$}}}}}
\put(60,120){\makebox(0,0)[lb]{\raisebox{0pt}[0pt][0pt]{\shortstack[l]{{\twlrm
$\zeta_V$}}}}}
\put(0,175){\makebox(0,0)[lb]{\raisebox{0pt}[0pt][0pt]{\shortstack[l]{{\twlrm
$\zeta_H$}}}}}
\end{picture}

where
\begin{eq}
g_a&=&{\Theta_{x^{2N}}(-x^{4a^*})h(1;0,0)\over
\Theta_{x^{2N}}(-1)h(x;0,0)},\quad g_\bullet=1,\\
h(\zeta; a, b)&=&f(\zeta; a - b) f(-\zeta; a + b),\nonumber\\
f(\zeta;a)&=&\prod_{l=0}^{a^*-1}\Theta_{x^{2N}}(\zeta^{-1}x^{2l+1})
\prod_{l=a^*}^{n-1}\Theta_{x^{2N}}(\zeta x^{2l+1}),\nonumber\\
\kappa(\zeta)&=&(x^{2N};x^{2N})_\infty^{2n}
\frac{(x^2\zeta^2;x^4,x^4)_\infty (x^4\zeta^{-2};x^4,x^4)_\infty}
{(x^4\zeta^2;x^4,x^4)_\infty (x^6\zeta^{-2};x^4,x^4)_\infty}
\times \hat{\kappa}(\zeta),\n
\hat{\kappa}(\zeta)&=&1\hskip 7cm  (N=2n), \n
&=&\frac{(x^{2N+2}\zeta^2;x^4,x^{4N})_\infty
(x^{2N+4}\zeta^{-2};x^4,x^{4N})_\infty}
{(x^{2N}\zeta^2;x^4,x^{4N})_\infty
(x^{2N+2}\zeta^{-2};x^4,x^{4N})_\infty}
\qquad(N=2n+1). \nonumber
\end{eq}

With these Boltzmann weights, we have (\ref{eqn:init})-(\ref{eqn:cross})
for the KM model. (In order to adjust the crossing symmetry to
the same formulation as in the ABF case, we have modified the
parametrization (3.3) in \cite{JMO}.)
The possible choices of the boundary conditions are similar to
(\ref{eqn:b.c.1}), (\ref{eqn:b.c.2}):
\begin{eq}
\l_{(j_1,j_2)}&=&0\quad \hbox{ if $j_1+j_2=i+1 \bmod2$,}\nonumber\\
&=&\bullet \quad \hbox{ if $j_1+j_2=i \bmod2$.}\nonumber
\end{eq}
(For technical reasons we restrict to the simplest case.)
Then all the consequences of these equations
are equally valid in the KM model, including in particular
the $q$-difference equations. Unlike the ABF models, however, we do not have
a representation theoretic picture of the KM models.
In the following section, restricting to the Ising case $N=2$, we give
a mathematical model of the VO's in terms of the free fermion algebra.

\section{Vertex operators in the Ising model}

The Ising model is the simplest
special case of both the ABF and the KM models.
In this section we wish to reexamine it in the framework
of this paper.
The special feature about the Ising model is that
we are interested in is that
one can diagonalize the CTM's using Jordan-Wigner fermions,
and obtain the VO's explicitly.
For the reader's convenience we shall repeat the formulation of VO's
in the context of the Ising model.

\subsection{Boltzmann weights}
We choose to work in terms of the edge formulation of the model.
In the literature, the Boltzmann weights are usually given in terms of
the anisotropic coupling constants $K$ and $L$ as

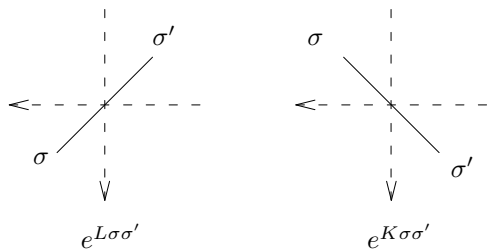
\begin{center}

\setlength{\unitlength}{0.0125in}
\begin{picture}(209,115)(0,-10)
\drawline(20,40)(60,80)
\dashline{4.000}(40,100)(40,20)
\drawline(38.000,28.000)(40.000,20.000)(42.000,28.000)
\dashline{4.000}(80,60)(0,60)
\drawline(8.000,62.000)(0.000,60.000)(8.000,58.000)
\dashline{4.000}(200,60)(120,60)
\drawline(128.000,62.000)(120.000,60.000)(128.000,58.000)
\dashline{4.000}(160,100)(160,20)
\drawline(158.000,28.000)(160.000,20.000)(162.000,28.000)
\drawline(180,40)(140,80)
\put(150,0){\makebox(0,0)[lb]{\raisebox{0pt}[0pt][0pt]{\shortstack[l]{{\twlrm
$e^{K\sigma\sigma'}$}}}}}
\put(30,0){\makebox(0,0)[lb]{\raisebox{0pt}[0pt][0pt]{\shortstack[l]{{\twlrm
$e^{L\sigma\sigma'}$}}}}}
\put(185,30){\makebox(0,0)[lb]{\raisebox{0pt}[0pt][0pt]{\shortstack[l]{{\twlrm
$\sigma'$}}}}}
\put(60,85){\makebox(0,0)[lb]{\raisebox{0pt}[0pt][0pt]{\shortstack[l]{{\twlrm
$\sigma'$}}}}}
\put(125,85){\makebox(0,0)[lb]{\raisebox{0pt}[0pt][0pt]{\shortstack[l]{{\twlrm
$\sigma$}}}}}
\put(10,35){\makebox(0,0)[lb]{\raisebox{0pt}[0pt][0pt]{\shortstack[l]{{\twlrm
$\sigma$}}}}}
\end{picture}
\end{center}

We work in the ferromagnetic low-temperature regime $K,L > 0$.
Following \cite{BaxBk}, let the coupling constants be parametrized as
\begin{eqnarray*}
&&\sinh(2K)= -i   {\rm sn}(iu),\quad \cosh(2K)={\rm cn}(iu),\\
&&\sinh(2L)=  i k^{-1}{\rm ns}(iu),\quad \cosh(2L)=ik^{-1} {\rm ds}(iu).
\end{eqnarray*}
where $u=u_V-u_H$ is the difference of (additive)
spectral parameters: $u_V$ attached to the vertical lines
and $u_H$ attached to the horizontal lines.
The $sn$ and $cn$ are the elliptic functions with half
periods $I$, $iI'$ corresponding to the modulus
$k=(\sinh 2K \sinh 2L)^{-1}$ (\cite{BaxBk}, Chapter 15).
We have also used Glashier's notation ${\rm ns}(u)=1/{\rm sn}(u)$,
${\rm ds}(u)={\rm dn}(u)/{\rm sn}(u)$, etc.
The region we consider is then $0<k<1$, $0 < u < I'$.
Notice that changing $u$ to $I'-u$
has the effect of exchanging $K$ and $L$ (crossing symmetry).

To compare this with the KM model for $N=2$ (see the last figure in Section 1),
let us write $w_\pm(\zeta)=\w(\bullet,b,\bullet,a,\zeta)$,
$\bw_\pm(\zeta)=\w(a,\bullet,b,\bullet,\zeta)$
($a,b=0,1$)
where $\pm$ is chosen according to whether $a=b$ or $a\neq b$.
Then we have
\begin{eqnarray}
w_+(\zeta)&=&
{1\over \tilde{\kappa}(\zeta)}
(x^2\zeta^2;x^8)_\infty(x^6\zeta^{-2};x^8)_\infty, \n
w_-(\zeta)&=&
{1\over \tilde{\kappa}(\zeta)}
\zeta\,(x^2\zeta^{-2};x^8)_\infty(x^6\zeta^2;x^8)_\infty, \n
\bw_{\pm}(\zeta)&=&{1\over \chi}w_\pm(x/\zeta), \n
\tilde{\kappa}(\zeta)&=&
{(x^2\zeta^2;x^4,x^4)_\infty \over (x^4\zeta^{2};x^4,x^4)_\infty}
{(x^4\zeta^{-2};x^4,x^4)_\infty \over (x^6\zeta^{-2};x^4,x^4)_\infty},\n
\chi&=&{(x^4;x^8)^2_\infty \over (x^2;x^4)_\infty}.
\nonumber
\end{eqnarray}
Identifying  $x = exp(- \pi I'/2I)$, $\zeta = exp(-\pi u/2I)$ we find
$w_-(\zeta)/ w_+(\zeta)= {\rm cn}\,(iu)+i{\rm sn}\,(iu)$.
Hence, up to the common scalar ${\tilde \kappa}(\zeta)$,
the KM parametrization agrees with the one for the coupling constants $K,L$.

\subsection{Corner transfer matrices}
There are two possible ways to divide the Ising lattice into 4 quadrants.
The choices depend on whether the site common to all quadrants
carries a state variable or does not.
Thus there are {\em two} CTM's to consider, and the complete space of states of
the model divides into two sectors.
This is analogous to the situation in the Ising CFT on the
annulus,
where the space of states divides into Neveu-Schwarz (NS) and Ramond (R)
sectors, that are half-integrally and integrally graded, respectively.
This will be made explicit below.
The possibilities are:

1. The center of the lattice carries a state variable. In this case,
the NW-quadrant CTM has the form

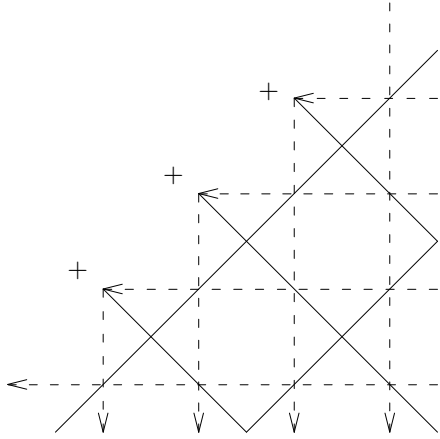
\begin{center}

\setlength{\unitlength}{0.0125in}
\begin{picture}(180,195)(0,-10)
\drawline(100,0)(40,60)
\drawline(180,80)(120,140)
\drawline(180,0)(80,100)
\drawline(180,80)(100,0)
\drawline(180,160)(20,0)
\dashline{4.000}(180,140)(120,140)
\drawline(128.000,142.000)(120.000,140.000)(128.000,138.000)
\dashline{4.000}(180,100)(80,100)
\drawline(88.000,102.000)(80.000,100.000)(88.000,98.000)
\dashline{4.000}(180,60)(40,60)
\drawline(48.000,62.000)(40.000,60.000)(48.000,58.000)
\dashline{4.000}(180,20)(0,20)
\drawline(8.000,22.000)(0.000,20.000)(8.000,18.000)
\dashline{4.000}(40,60)(40,0)
\drawline(38.000,8.000)(40.000,0.000)(42.000,8.000)
\dashline{4.000}(80,100)(80,0)
\drawline(78.000,8.000)(80.000,0.000)(82.000,8.000)
\dashline{4.000}(120,140)(120,0)
\drawline(118.000,8.000)(120.000,0.000)(122.000,8.000)
\dashline{4.000}(160,180)(160,0)
\drawline(158.000,8.000)(160.000,0.000)(162.000,8.000)
\put(25,65){\makebox(0,0)[lb]{\raisebox{0pt}[0pt][0pt]{\shortstack[l]{{\twlrm
$+$}}}}}
\put(65,105){\makebox(0,0)[lb]{\raisebox{0pt}[0pt][0pt]{\shortstack[l]{{\twlrm
$+$}}}}}
\put(105,140){\makebox(0,0)[lb]{\raisebox{0pt}[0pt][0pt]{\shortstack[l]{{\twlrm
$+$}}}}}
\end{picture}
\end{center}

Since the state variable at the center can be either $+1$ or $-1$,
this sector further subdivides into two sub-sectors. Once again, this
subdivision is analogous to the situation in conformal field theory,
where the NS sector subdivides into two subsectors.

For that reason, we refer to this sector as the NS sector and write
the space of states as $\H^\rN=\H^{\rN+}\oplus \H^{\rN -}$,
with the $\pm$ referring to the mentioned subsectors.
The corresponding CTM will be denoted by $A_4^{\rN}(\zeta)=\zeta^{D^{\rN}}$.
Here $D^{\rN}$  is the CTM Hamiltonian, that the YB equations
guarantee to be independent of $\zeta$.
The subsectors $\H^{\rN,\pm}$ are invariant under the action of $D^{\rN}$.

2. The center of the lattice is not on the physical lattice.
In this case the NW-quadrant is

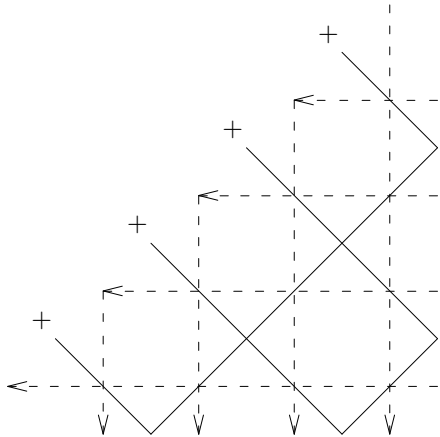
\begin{center}

\setlength{\unitlength}{0.0125in}
\begin{picture}(180,195)(0,-10)
\drawline(180,40)(140,0)
\drawline(180,120)(60,0)
\drawline(60,0)(20,40)
\drawline(140,0)(60,80)
\drawline(180,120)(140,160)
\drawline(180,40)(100,120)
\dashline{4.000}(180,140)(120,140)
\drawline(128.000,142.000)(120.000,140.000)(128.000,138.000)
\dashline{4.000}(180,100)(80,100)
\drawline(88.000,102.000)(80.000,100.000)(88.000,98.000)
\dashline{4.000}(180,60)(40,60)
\drawline(48.000,62.000)(40.000,60.000)(48.000,58.000)
\dashline{4.000}(180,20)(0,20)
\drawline(8.000,22.000)(0.000,20.000)(8.000,18.000)
\dashline{4.000}(40,60)(40,0)
\drawline(38.000,8.000)(40.000,0.000)(42.000,8.000)
\dashline{4.000}(80,100)(80,0)
\drawline(78.000,8.000)(80.000,0.000)(82.000,8.000)
\dashline{4.000}(120,140)(120,0)
\drawline(118.000,8.000)(120.000,0.000)(122.000,8.000)
\dashline{4.000}(160,180)(160,0)
\drawline(158.000,8.000)(160.000,0.000)(162.000,8.000)
\put(10,45){\makebox(0,0)[lb]{\raisebox{0pt}[0pt][0pt]{\shortstack[l]{{\twlrm
$+$}}}}}
\put(50,85){\makebox(0,0)[lb]{\raisebox{0pt}[0pt][0pt]{\shortstack[l]{{\twlrm
$+$}}}}}
\put(90,125){\makebox(0,0)[lb]{\raisebox{0pt}[0pt][0pt]{\shortstack[l]{{\twlrm
$+$}}}}}
\put(130,165){\makebox(0,0)[lb]{\raisebox{0pt}[0pt][0pt]{\shortstack[l]{{\twlrm
$+$}}}}}
\end{picture}
\end{center}

Once again, by analogy with conformal field theory, we will refer to
this sector as $\H^\rR$ and the CTM as $A_4^\rR(\zeta)=\zeta^{D^\rR}$.

The CTM's for the rest of the quadrants can be obtained from the above by
crossing symmetry.
To be precise,
in the definition of the CTM's we have to make a definite choice regarding
the boundary conditions at infinity. We choose the $+$ boundary conditions.

\subsection{Diagonalization of CTM}
The starting point of the diagonalization procedure
is to formulate the model in terms of Jordan-Wigner
(JW) fermions:
\begin{eqnarray*}
&&\psn_0=\sigma^z_1,\quad
\psn_1=\sigma^y_1,\quad
\psn_{2j}=\sigma^z_{j+1}\sigma^x_j\cdots \sigma^x_1, \quad
\psn_{2j+1}=\sigma^y_{j+1}\sigma^x_j\cdots \sigma^x_1,  \\
&&\psr_1=\sigma^z_1,\quad
\psr_2=\sigma^y_1,\quad
\psr_{2j-1}=\sigma^z_j\sigma^x_{j-1}\ldots \sigma^x_1, \quad
\psr_{2j}=\sigma^y_j\sigma^x_{j-1}\ldots \sigma^x_1.
\end{eqnarray*}
Here the indices of the Pauli matrices
refer to the lattice sites labeled by $1,2,\cdots$ from the central site
upwards or to the left.
They satisfy
\[
[\psi^\rN_j,\psi^\rN_k]_+=2\delta_{jk},\quad
[\psi^\rR_j,\psi^\rR_k]_+=2\delta_{jk}.
\]

Expressed in terms of the JW fermions, the CTM Hamiltonians are
\begin{eqnarray*}
D^{\rN}&=&-{iI\over \pi}
\bigl(\sum_{j\ge0}(2j+1)\psi^\rN_{2j+1}\psi^\rN_{2j+2}
+k\sum_{j\ge1}2j\psi^\rN_{2j}\psi^\rN_{2j+1}\bigr),\\
D^{\rR}&=&-{iI\over \pi}\bigl(
k\sum_{j\ge0}(2j+1)\psi^\rR_{2j+1}\psi^\rR_{2j+2}
+\sum_{j\ge1}2j\psi^\rR_{2j}\psi^\rR_{2j+1}\bigr)
\end{eqnarray*}
Since these are quadratic in the fermions, it is in principle straightforward
to diagonalize them.
Namely we would like to find a set of fermions
\begin{eqnarray*}
&&\phi^\rN_r=\sum_{j=1}^\infty a_{rj}^\rN\psi_j^\rN,\quad
  \phi^\rR_r=\sum_{j=1}^\infty a_{rj}^\rR\psi_j^\rR,\quad
\end{eqnarray*}
that diagonalize the adjonit action of the CTM Hamiltonians
\begin{eqnarray*}
&&[D^\rN,\phi^\rN_r]=-2r\phi^\rN_r,\quad
[D^\rR,\phi^\rR_r]=-2r\phi^\rR_r.
\end{eqnarray*}
As it turns out the eigenvalues are labeled by
$r\in\Zh$ for the $NS$ sector and by $r\in\Z$ for the $R$ sector.
We shall also require that these fermions obey the anti-commutation relations
\begin{eqnarray}
&&[\phi^\rN_r,\phi^\rN_s]_+=\eta_r\delta_{r+s,0},\
[\phi^\rR_r,\phi^\rR_s]_+=\eta_r\delta_{r+s,0},\ \eta_r=x^{2r}+x^{-2r},
\label{eqn:creann}
\end{eqnarray}
where the normalization factor $\eta_r$ is put in for convenience.
Conversely the JW fermions can be expressed as
\begin{eqnarray*}
&&\psi_j^\rN=\sum_{r}2\eta_r^{-1} a_{-r\,j}^\rN\phi^\rN_r,\quad
\psi_j^\rR=\sum_{r}2\eta_r^{-1} a_{-r\,j}^\rR\phi^\rR_r.
\end{eqnarray*}

The determination of  $a_{rj}^\rN$ and $a_{rj}^\rR$
is described in \cite{BaxCTM}, \cite{Dav}.
Substituting the above expansions in the commutation relations
and using the expression of the Hamiltonians in terms of JW fermions,
we obtain second order linear
differential equations for the generating functions
\begin{eqnarray*}
a^{\rN}_{r}(t)=\sum_{j=1}^\infty a^{\rN}_{rj}t^j,\quad
a^{\rR}_{r}(t)=\sum_{j=1}^\infty a^{\rR}_{rj}t^j.
\end{eqnarray*}
For (\ref{eqn:creann}) to make sense we demand that the coefficients
$a_{rj}^\rN$ and $a_{rj}^\rR$ be square summable.
Changing variables by $t=\sqrt{k}\,{\rm sn}(v)$ and
solving the equations under this regularity condition we obtain
\begin{eqnarray}
&&a^{\rm NS}_{r}(t)=\sqrt{\pi \over2\bK}\left(
{\rm sc}(v)\cos\bigl({\pi r\over\bK} v\bigr)
-i\sqrt{k}\,{\rm sd}(v)\sin\bigl({\pi r\over\bK} v\bigr)\right),
\label{eqn:aNS}\\
&&a^{\rm R}_{r}(t)=\sqrt{\pi \over2\bK}\left(
\sqrt{k}\,{\rm sd}(v)\cos\bigl({\pi r\over\bK} v\bigr)
-i{\rm sc}(v)\sin\bigl({\pi r\over\bK} v\bigr)\right).
\label{eqn:aR}
\end{eqnarray}
They are single-valued and holomorphic for $|t|<1/\sqrt{k}$.

\subsection{The Fock spaces}
The `Hilbert spaces' of states $\Hn$ and $\Hr$ are the Fock spaces
of $\phi^\rN_r$ and $\phi^\rR_r$ ($r\ne 0$)
over the vacuum states $\vacn$ and $\vacr$, respectively.
The dual Fock spaces have dual vacuum states $\dvacn$ and
$\dvacr$, normalized as usual.

The vacuum states satisfy the usual creation and annihilation conditions with
respect to $\phi^\rN_r$ and $\phi^\rR_r$:
\begin{eq}
&\dvacn\phi^\rN_r=0\quad (r<0) \qquad
&\phi^\rN_r\vacn=0\quad (r>0),\n
&\dvacr\phi^\rR_r=0\quad (r<0),\qquad
&\phi^\rR_r\vacr=0\quad (r>0).
\end{eq}
The CTM Hamiltonians are to be normalized so that
\begin{eqnarray*}
&&D^\rN\vacn=D^\rR\vacr=\dvacn D^\rN=\dvacr D^\rR=0.
\end{eqnarray*}
They give a grading of the spaces
$\H^\rN$, $\H^\rR$ by non-negative integers:
\begin{eqnarray*}
&&\H^{\rN,+}=\bigoplus_{d\in\Z_{\ge0},d:even}\Hn_{d},\quad
\H^{\rN,-}=\bigoplus_{d\in \Z_{\ge0},d:odd}\Hn_{d},\\
&&\Hr=\bigoplus_{d\in\Z\ge0,d:even}\Hr_{d},
\end{eqnarray*}
where $\Hn_d=\{v\in\Hn \mid D^{\rN}v=d\,v \}$ and likewise for $\Hr_d$.

Besides the creation/annihilation operators, the fermion
algebra in each sector contains an extra one
 $\psi_0^\rN$, $\phi^\rR_0$:
they
commute with the CTM's, anticommute with the creation-annihilation fermions,
and satisfy $\bigl(\psi_0^{\rN}\bigr)^2=1$,
$\bigl(\phi^{\rR}_0\bigr)^2=1$.
Hence their action should fix the vacuum vector up to sign.
The choice of the $+$ boundary condition corresponds to
\begin{eqnarray*}
&\psi^\rN_0\vacn=\vacn,\quad &\dvacn\psi^\rN_0=\dvacn,\\
&\phi^\rR_0\vacr=\vacr,\quad &\dvacr\phi^\rR_0=\dvacr.
\end{eqnarray*}
In particular $\H^{\rN,\pm}$ is the eigenspace of $\psi^\rN_0$ with
the eigenvalue $\pm 1$.

\subsection{Vertex operators}
As for CTM's, there are {\em two} types of VO's depending on which sector
they act on:
\begin{eqnarray*}
&&\Phi_{\rN}^{\rR\, {\rm V}}(\zeta):\H^{\rN}\goto{} \Hr, \\
&&\Phi_{\rR}^{\rN\, {\rm V}}(\zeta):\Hr\goto{} \H^{\rN}.
\end{eqnarray*}
Here the superscript V indicates that the VO is `vertical'.
Sometimes the NS-subsectors are indicated by $\sigma=\pm$; thus
$\Phi^{\rR V}_{\rN\sigma}(\zeta)=
\Phi^{\rR V}_{\rN}(\zeta)\bigr|_{\H^{\rN\sigma}}$,
$\Phi_{\rR }^{\rN,\sigma,V}(\zeta)=
P^\sigma\Phi_{\rR }^{\rN V}(\zeta)$ where
$P^\sigma$ denotes the projection onto $\H^{\rN\sigma}$.

There are also the `horizontal' versions of VO's
$\Phi_{\rN\,\sigma}^{\rR\, {\rm H}}(\zeta)$,
$\Phi_{\rR}^{\rN\,\sigma {\rm H}}(\zeta)$.
The following figures show the $({\bf s},{\bf s'})$-elements of these VO's
where ${\bf s}=(s_1,s_2,\ldots)$ and so on. The NS-subsectors are indicated
by $\sigma$.

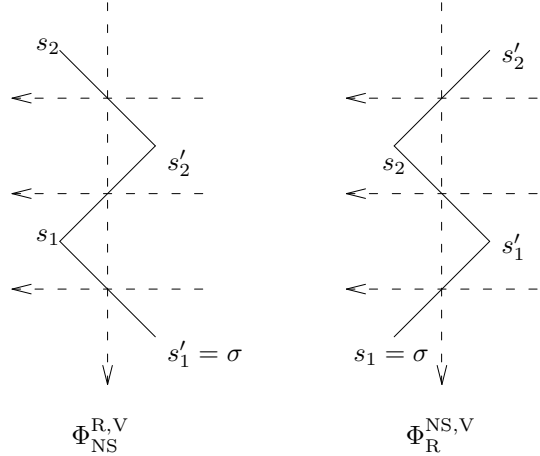
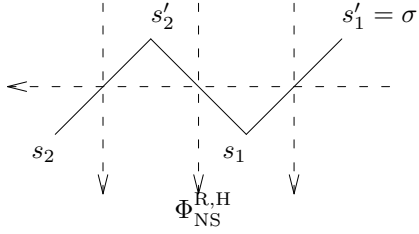
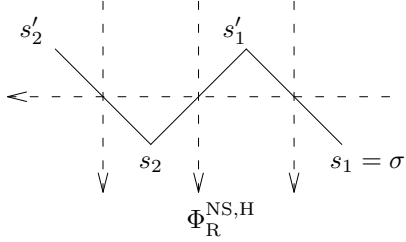
\begin{center}

\setlength{\unitlength}{0.0125in}
\begin{picture}(241,200)(0,-10)
\drawline(200,165)(160,125)(200,85)(160,45)
\drawline(20,165)(60,125)(20,85)(60,45)
\dashline{4.000}(40,185)(40,25)
\drawline(38.000,33.000)(40.000,25.000)(42.000,33.000)
\dashline{4.000}(80,65)(0,65)
\drawline(8.000,67.000)(0.000,65.000)(8.000,63.000)
\dashline{4.000}(80,105)(0,105)
\drawline(8.000,107.000)(0.000,105.000)(8.000,103.000)
\dashline{4.000}(80,145)(0,145)
\drawline(8.000,147.000)(0.000,145.000)(8.000,143.000)
\dashline{4.000}(220,145)(140,145)
\drawline(148.000,147.000)(140.000,145.000)(148.000,143.000)
\dashline{4.000}(220,105)(140,105)
\drawline(148.000,107.000)(140.000,105.000)(148.000,103.000)
\dashline{4.000}(220,65)(140,65)
\drawline(148.000,67.000)(140.000,65.000)(148.000,63.000)
\dashline{4.000}(180,185)(180,25)
\drawline(178.000,33.000)(180.000,25.000)(182.000,33.000)
\put(165,0){\makebox(0,0)[lb]{\raisebox{0pt}[0pt][0pt]{\shortstack[l]{{\twlrm
$\Phi^{\rm NS,V}_{\rm R}$}}}}}
\put(25,0){\makebox(0,0)[lb]{\raisebox{0pt}[0pt][0pt]{\shortstack[l]{{\twlrm
$\Phi^{\rm R,V}_{\rm NS}$}}}}}
\put(155,115){\makebox(0,0)[lb]{\raisebox{0pt}[0pt][0pt]{\shortstack[l]{{\twlrm
$s_2$}}}}}
\put(143,35){\makebox(0,0)[lb]{\raisebox{0pt}[0pt][0pt]{\shortstack[l]{{\twlrm
$s_1=\sigma$}}}}}
\put(205,160){\makebox(0,0)[lb]{\raisebox{0pt}[0pt][0pt]{\shortstack[l]{{\twlrm
$s'_2$}}}}}
\put(205,80){\makebox(0,0)[lb]{\raisebox{0pt}[0pt][0pt]{\shortstack[l]{{\twlrm
$s'_1$}}}}}
\put(65,115){\makebox(0,0)[lb]{\raisebox{0pt}[0pt][0pt]{\shortstack[l]{{\twlrm
$s'_2$}}}}}
\put(65,35){\makebox(0,0)[lb]{\raisebox{0pt}[0pt][0pt]{\shortstack[l]{{\twlrm
$s'_1=\sigma$}}}}}
\put(10,165){\makebox(0,0)[lb]{\raisebox{0pt}[0pt][0pt]{\shortstack[l]{{\twlrm
$s_2$}}}}}
\put(10,85){\makebox(0,0)[lb]{\raisebox{0pt}[0pt][0pt]{\shortstack[l]{{\twlrm
$s_1$}}}}}
\end{picture}

\setlength{\unitlength}{0.0125in}
\begin{picture}(171,251)(0,-10)
\dashline{4.000}(40,95)(40,15)
\drawline(38.000,23.000)(40.000,15.000)(42.000,23.000)
\dashline{4.000}(80,230)(80,150)
\drawline(78.000,158.000)(80.000,150.000)(82.000,158.000)
\drawline(140,35)(100,75)(60,35)(20,75)
\drawline(140,215)(100,175)(60,215)(20,175)
\dashline{4.000}(160,195)(0,195)
\drawline(8.000,197.000)(0.000,195.000)(8.000,193.000)
\dashline{4.000}(40,230)(40,150)
\drawline(38.000,158.000)(40.000,150.000)(42.000,158.000)
\dashline{4.000}(120,230)(120,150)
\drawline(118.000,158.000)(120.000,150.000)(122.000,158.000)
\dashline{4.000}(120,95)(120,15)
\drawline(118.000,23.000)(120.000,15.000)(122.000,23.000)
\dashline{4.000}(80,95)(80,15)
\drawline(78.000,23.000)(80.000,15.000)(82.000,23.000)
\dashline{4.000}(160,55)(0,55)
\drawline(8.000,57.000)(0.000,55.000)(8.000,53.000)
\put(75,0){\makebox(0,0)[lb]{\raisebox{0pt}[0pt][0pt]{\shortstack[l]{{\twlrm
$\Phi^{\rm NS,H}_{\rm R}$}}}}}
\put(70,140){\makebox(0,0)[lb]{\raisebox{0pt}[0pt][0pt]{\shortstack[l]{{\twlrm
$\Phi^{\rm R,H}_{\rm NS}$}}}}}
\put(90,80){\makebox(0,0)[lb]{\raisebox{0pt}[0pt][0pt]{\shortstack[l]{{\twlrm
$s'_1$}}}}}
\put(5,80){\makebox(0,0)[lb]{\raisebox{0pt}[0pt][0pt]{\shortstack[l]{{\twlrm
$s'_2$}}}}}
\put(135,25){\makebox(0,0)[lb]{\raisebox{0pt}[0pt][0pt]{\shortstack[l]{{\twlrm
$s_1=\sigma$}}}}}
\put(55,25){\makebox(0,0)[lb]{\raisebox{0pt}[0pt][0pt]{\shortstack[l]{{\twlrm
$s_2$}}}}}
\put(90,165){\makebox(0,0)[lb]{\raisebox{0pt}[0pt][0pt]{\shortstack[l]{{\twlrm
$s_1$}}}}}
\put(10,165){\makebox(0,0)[lb]{\raisebox{0pt}[0pt][0pt]{\shortstack[l]{{\twlrm
$s_2$}}}}}
\put(140,222){\makebox(0,0)[lb]{\raisebox{0pt}[0pt][0pt]{\shortstack[l]{{\twlrm
$s'_1=\sigma$}}}}}
\put(60,222){\makebox(0,0)[lb]{\raisebox{0pt}[0pt][0pt]{\shortstack[l]{{\twlrm
$s'_2$}}}}}
\end{picture}
\end{center}

The crossing symmetry implies
\begin{eqnarray*}
&&\Phi^{\rR,{\rm H}}_{\rN,\sigma}(\zeta)=
\Phi^{\rR,{\rm V}}_{\rN,\sigma}(x/\zeta),\quad
\Phi^{\rN,\sigma,{\rm H}}_\rR(\zeta)=
\Phi^{\rN,\sigma,{\rm V}}_\rR(x/\zeta).
\end{eqnarray*}
Arguing similarly as in the previous section
(the derivation of (\ref{eqn:Facehomo})),
we see that the $\zeta$-dependence of the VO's is given by
\begin{eq}
\Phi^{\rR,{\rm V}}_{\rN,\sigma}(\zeta)&=&
\zeta^{-D^\rR}\Phi^{\rR,{\rm V}}_{\rN,\sigma}(1)\zeta^{D^\rN}, \\
\Phi^{\rN,\sigma,{\rm V}}_{\rR}(\zeta)&=&
\zeta^{-D^\rN}\Phi^{\rN,\sigma,{\rm V}}_{\rR}(1)\zeta^{D^\rR}.
\end{eq}
We will adopt the  normalization
\begin{eqnarray}
&&\dvacr\Phi^{\rR,{\rm V}}_{\rN,+}(1)\vacn=1,\quad
\dvacn\Phi^{\rN,+,{\rm V}}_{\rR}(1)\vacr=1. \label{eqn:normVO}
\end{eqnarray}

\subsection{Intertwining Property}
The Boltzmann weights enjoy the initial condition
\[
w_\pm(1)=1, \qquad \bw_+(1)=1,\quad \bw_-(1)=0.
\]
Hence at $\zeta=1$ the VO's become particularly simple:
e.g. $\Phi^{\rN,{\rm V}}_{\rR}(1)$
carries a vertical spin configuration to an identical one on the
next left column.
However it is {\em not} the identity because the domain $\H^\rR$
and the range $\H^\rN$ are different spaces.
Using the JW fermions in each sector,
the preservation of spins can be
equivalently stated as the intertwining property
\begin{eq}
\psi^\rN_j\Phi^{\rN,\sigma,{\rm V}}_\rR(1)
&=&\Phi^{\rN,-\sigma,{\rm V}}_\rR(1)\psi^\rR_{j+1}\quad(j\ge1),
\label{eqn:RNV}\\
\sigma \Phi^{\rN,\sigma,{\rm V}}_\rR(1)&=&
\Phi^{\rN,\sigma,{\rm V}}_\rR(1)\psi^\rR_1.
\label{eqn:RNV0}
\end{eq}
In the same manner we find the following for the other types of VO's:
\begin{eqnarray*}
\psi^\rR_j\Phi^{\rR,{\rm V}}_{\rN,\sigma}(1)
&=&\Phi^{\rR,{\rm V}}_{\rN,-\sigma}(1)\psi^\rN_{j+1}\quad(j\ge1),
\label{eqn:NRV}\\
i\sigma\Phi^{\rR,{\rm V}}_{\rN,\sigma}(1)&=&
\Phi^{\rR,{\rm V}}_{\rN,-\sigma}(1)\psi^\rN_1,\label{eqn:NRV0}
\end{eqnarray*}
\begin{eqnarray*}
\psi^\rN_{j+1}\Phi^{\rN,\sigma,{\rm H}}_\rR(1)
&=&\Phi^{\rN,-\sigma,{\rm H}}_\rR(1)\psi^\rR_j\quad(j\ge1),
\label{eqn:RNH}\\
\psi^\rN_1\Phi^{\rN,-\sigma,{\rm H}}_\rR(1)&=&
-i\sigma\Phi^{\rN,\sigma,{\rm H}}_\rR(1),
\label{eqn:RNH0}
\end{eqnarray*}
\begin{eqnarray*}
\psi^\rR_{j+1}\Phi^{\rR,{\rm H}}_{\rN,\sigma}(1)
&=&\Phi^{\rR,{\rm H}}_{\rN,-\sigma}(1)\psi^\rN_j\quad(j\ge1),
\label{eqn:NRH}\\
\psi^\rR_1\Phi^{\rR,{\rm H}}_{\rN,\sigma}(1)&=&
\sigma\Phi^{\rR,{\rm H}}_{\rN,\sigma}(1).
\label{eqn:NRH0}
\end{eqnarray*}
Along with the normalization condition (\ref{eqn:normVO}),
these commutation relations with the fermions
characterize the VO's uniquely, and all their matrix elements can be
determined.
The working is described in Appendix A.
We obtain the following formulas for the general matrix elements:
\begin{eqnarray*}
\dvacr \pr_{l_1}\cdots\pr_{l_n}\Phi^{\rR\,V}_{\rN}(\zeta)
\pn_{-k_1}\cdots\pn_{-k_m}
\vacn /\zeta^{2(-l_1-\ldots-l_n+k_1+\ldots+k_m)}\\
= \prod x^{2l_j}\gamma_{l_j} \prod \left(\sqrt{-1} x^{-1/2}\,
\gamma_{k_i-1/2}\right)
\times \prod_{j<j'}X_{l_j,l_{j'}}
\prod_{i,j}X_{l_j,-k_i}\prod_{i<i'}X_{k_i,k_{i'}},
\end{eqnarray*}
\begin{equation}
\label{eqn:matele1}
\end{equation}
\begin{eqnarray*}
\dvacn \pn_{k_m}\cdots\pn_{k_1}\Phi^{\rN\,V}_\rR(\zeta)
\pr_{-l_n}\cdots\pr_{-l_1} \vacn
/\zeta^{2(l_1+\ldots+l_n-k_1-\ldots-k_m)}\\
= \prod \gamma_{l_j} \prod \left(- \sqrt{-1}\,x^{2k_i-1/2}
\gamma_{k_i-1/2}\right)
\times \prod_{j<j'}X_{l_j,l_{j'}}
\prod_{i,j}X_{l_j,- k_i}\prod_{i<i'}X_{k_i,k_{i'}}.
\end{eqnarray*}
\begin{equation}
\label{eqn:matele2}
\end{equation}
Here we have set
\begin{eqnarray}
&&X_{k,l}={x^{4k}-x^{4l} \over 1-x^{4(k+l)}}=-X_{l,k},
\quad X_{-k,-l}=X_{k,l},\quad
\gamma_n={(x^2;x^4)_n \over (x^4;x^4)_n}. \label{eqn:xgamma}
\end{eqnarray}

\subsection{Vacuum-to-vacuum two point functions}
In what follows,
we shall use the shorthand notations
\[
\Phi^{\rN,\sigma,V}_\rR(\zeta)=\Phi^\sigma(\zeta),\quad
\Phi_{\rN,\sigma}^{\rR,V}(\zeta)=\Phi_\sigma(\zeta).
\]
When no confusion may arise,
we shall also drop $NS$, $R$ and write $\vac$ for
either $\vacn$ or $\vacr$, etc.

Using the formulas for the matrix elements
(\ref{eqn:matele1})-(\ref{eqn:matele2}) one can in principle
compute the vacuum expectation values of products of VO's.
The expansion in powers of $x$ suggests that the two point functions
are given by the following infinite products:
setting $\zeta=\zeta_1/\zeta_2$ we have
\begin{eqnarray}
\sum_\sigma \dvac \Phi_{\sigma}(\zeta_1)
\Phi^{\sigma}(\zeta_2) \vac
&=&{(x^3\zeta;x^4)_\infty \over (x\zeta;x^4)_\infty }F_+(\zeta^2),
\label{eqn:vacexp1}\\
\dvac \Phi^+(\zeta_1)\Phi_+(\zeta_2) \vac& =&F_-(\zeta^2),
\label{eqn:vacexp2}\\
\dvac \phi^{\rm NS}_{1/2}\Phi^-(\zeta_1)\Phi_-(\zeta_2)\phi^{\rm NS}_{-1/2}
 \vac &=&(\zeta^{-1}+\zeta)F_-(\zeta^2),
\label{eqn:vacexp3}\\
\dvac \Phi^+(\zeta_1)\Phi_-(\zeta_2)\phi^{\rm NS}_{-1/2} \vac &=&
ix^{-1/2}\zeta_2 F_+(\zeta^2),
\label{eqn:vacexp4}\\
\dvac \phi^{\rm NS}_{1/2}\Phi^-(\zeta_1)\Phi_+(\zeta_2) \vac
&=&-ix^{1/2}\zeta_1^{-1} F_+(\zeta^2).
\label{eqn:vacexp5}
\end{eqnarray}
Here
\begin{eqnarray*}
&&F_+(z)={(x^2z;x^4,x^8)_\infty^2\over
(x^2z;x^4)_\infty(x^4z;x^4,x^4)_\infty},\\
&&F_-(z)={(x^6z;x^4,x^8)_\infty^2\over
(x^4z;x^4,x^4)_\infty}.
\end{eqnarray*}
We shall derive these equations using the formulation of the
Ising model as a special case of the ABF models (see the end
of this section).

\noindent{\sl Remark.}\quad
In the language of the string theory, the vertex operator
$\Phi^{\rR\,V}_{\rN}(z^{-1/2})$,
for example, is expressed as
\begin{eqnarray*}
&&\Phi^\rR_{\rN}(z^{-1/2})=\dvacr :\exp Y:\vacn,\\
&&Y=-{1\over 2}\sum_{m,n>0}X_{m,n}
\varphi^\rR_{-m}\varphi^\rR_{-n}z^{m+n}
+\sum_{m,r>0}X_{m,-r}\varphi^\rR_{-m}\varphi^{\rN}_r z^{m-r}\\
&&-\phi^\rR_0(\sum_{m>0}\varphi^\rR_{-m}z^m
-\sum_{r>0}\varphi^{\rN}_r z^{-r})
-{1\over 2}\sum_{r,s>0}X_{r,s}
\varphi^{\rN}_{r}\varphi^{\rN}_{s}z^{-r-s},
\end{eqnarray*}
where we have set,
for $m, n \in {\bf Z}$ and $r, s \in {\bf Z}+{1\over 2}$,
\[
\varphi^\rR_{-m}=x^{2m}{\gamma_m\over \eta_m}\phi^\rR_{-m},\quad
\varphi^{\rN}_r=\sqrt{-1}x^{-1/2}
{\gamma_{r-{1\over2}}\over \eta_r}\phi^{\rN}_r.
\]
For the notation $\dvacr :\exp Y:\vacn$,
see the literature on the fermion emission vertex,
for example \cite{Scherk}.
For $x=1$, the vertex operators  $\Phi^\sigma(z^{-1/2})$ and
$\Phi_\sigma(z^{-1/2})$ are the primary fields  with the confomal weight 1/16
of the $c=1/2$ minimal model, up to some power of $z$.
It can be verified, by noting the identity below,
that (\ref{eqn:vacexp1})
and
(\ref{eqn:vacexp2})
correspond
to the four and two point functions of spin fields respectively \cite{BPZ}:
\begin{eqnarray*}
&&{(x^3\zeta;x^4)_\infty\over (x\zeta;x^4)_\infty}
{(x^2\zeta^2;x^8)_\infty\over (x^6\zeta^2;x^8)_\infty}\\
&&=\phi\left({{1/4~-1/4}\atop{1/2}};x^8,x^6\zeta^2\right)
+{x\zeta\over 1+x^2}
\phi\left({{1/4~3/4}\atop {3/2}};x^8,x^6\zeta^2\right),
\end{eqnarray*}
where
\begin{eqnarray*}
\phi\left({{a~b}\atop{c}};q,z\right)
=\sum_{n=0}^\infty{(q^a;q)_n(q^b;q)_n \over (q^c;q)_n(q;q)_n}z^n,
\quad (z;q)_n=\prod_{j=0}^{n-1}(1-zq^j),
\end{eqnarray*}
denotes the basic hypergeometric series.

\subsection{Unitarity and commutation relations}
We have the following relations:
\begin{eqnarray}
&&\sum_\sigma
\Phi_{\sigma}(x\zeta)
\Phi^\sigma(\zeta)=g^{\rm R}\times {\rm id}_{\H^\rR},
\label{eqn:inverse1}\\
&&
\Phi^{\sigma}(x\zeta)\Phi_{\sigma}(\zeta)
=g^{\rm NS}\times {\rm id}_{\H^{\rN,\sigma}}.
\label{eqn:inverse2}
\end{eqnarray}
The reason is that, from the intertwining properties of the VO's, the LHS
can be shown to commute
with the fermions. Since our Fock spaces are irreducible, they must then
be acting as scalars. The scalars can be determined by
considering the
vacuum-to-vacuum expectation values given above. In this way we get
\begin{eqnarray*}
&&g^{\rm R}=
{(x^4;x^4,x^8)_\infty^2 \over (x^2;x^4,x^4)_\infty},\qquad
g^{\rm NS}=
{(x^8;x^4,x^8)_\infty^2 \over (x^6;x^4,x^4)_\infty},\\
&&{g^{\rm R} \over g^{\rm NS}}=
{\tr_{\Hn}\left(x^{2D}\right) \over \tr_{\Hr}\left(x^{2D}\right)}
={(-x^2;x^4)_\infty \over (-x^4;x^4)_\infty}.
\end{eqnarray*}

Using a similar reasoning we can derive from (\ref{eqn:vacexp1})--
(\ref{eqn:vacexp5}) the following commutation relations:
\begin{eqnarray}
&&\Phi^{\sigma}(\zeta_2)
\Phi_{\sigma'}(\zeta_1)
=
\Phi^{\sigma}(\zeta_1)
\Phi_{\sigma'}(\zeta_2)
w_{\sigma \sigma'}(\zeta_1/\zeta_2), \label{eqn:comm1}\\
&&
\Phi_{\sigma}(\zeta_2)
\Phi^{\sigma}(\zeta_1)
=\sum_{\sigma'}
\Phi_{\sigma'}(\zeta_1)
\Phi^{\sigma'}(\zeta_2)
\bw_{\sigma \sigma'}(\zeta_1/\zeta_2).
\label{eqn:comm2}
\end{eqnarray}
which are what we expect.
\subsection{Difference equation}

Put
\begin{eqnarray*}
&&G_\pm^{\rm R}(\zeta_1/\zeta_2)
=\tr_{\Hr}\left(x^{2D}
\bigl(\Phi_{+}(\zeta_1)\Phi^{+}(\zeta_2)
\pm \Phi_{-}(\zeta_1)\Phi^{-}(\zeta_2)\bigr)
\right)
=G_+^{\rm R}(\pm\zeta_1/\zeta_2), \\
&&G_\pm^{\rm NS}(\zeta_1/\zeta_2)
=\tr_{\Hn}\left(x^{2D}
\bigl(\Phi^{+}(\zeta_1)\Phi_{+}(\zeta_2)
\pm \Phi^{-}(\zeta_1)\Phi_{-}(\zeta_2)\bigr)
\right)
=G_+^{\rm NS}(\pm \zeta_1/\zeta_2).
\end{eqnarray*}
{}From the commutation relations we then deduce that
\begin{eqnarray*}
&&w_+(\zeta)G^{\rm NS}_\pm(\zeta)=G^{\rm R}_\pm(x^2\zeta),\\
&&G_\pm
^{\rm R}(x^4\zeta)=w_+(x^2\zeta)\left(\bw_+(\zeta)\pm \bw_-(\zeta)\right)
G_\pm^{\rm R}(\zeta).
\end{eqnarray*}
The coefficient in the second equation can be factorized:
\begin{eq}
\bw_+(\zeta)\pm \bw_-(\zeta)
&=&
{(\pm x\zeta;x^4)_\infty \over (\pm x^3\zeta;x^4)_\infty}
{(\pm x^3\zeta^{-1};x^4)_\infty \over (\pm x\zeta^{-1};x^4)_\infty}
\nonumber\\
&&
\times
{(x^2\zeta^2;x^4)_\infty \over (x^2\zeta^{-2};x^4)_\infty}
{(x^4\zeta^2;x^4,x^4)_\infty \over (x^4\zeta^{-2};x^4,x^4)_\infty}
{(x^2\zeta^{-2};x^4,x^8)^2_\infty \over (x^2\zeta^2;x^4,x^8)_\infty^2 }.
\nonumber
\end{eq}
Using this we find the following solution:
\begin{eqnarray*}
&&G_\pm^{\rm R}(\zeta)=C
{(\pm x^3\zeta;x^4,x^4)_\infty \over
(\pm x\zeta;x^4,x^4)_\infty}
{(\pm x^7\zeta^{-1};x^4,x^4)_\infty \over
(\pm x^5\zeta^{-1};x^4,x^4)_\infty}
{f(\zeta) \over f(x)}, \\
&&
f(\zeta)=
{(x^2\zeta^2;x^4,x^8,x^8)^4_\infty \over (x^4\zeta^2;x^4,x^4,x^8)^2_\infty}
{(x^{10}\zeta^{-2};x^4,x^8,x^8)^4_\infty \over
(x^{12}\zeta^{-2};x^4,x^4,x^8)^2_\infty}
{(x^4\zeta^2;x^4,x^8)_\infty \over (x^2\zeta^2;x^4,x^8)^3_\infty
(x^8\zeta^{-2};x^2,x^8)_\infty}.
\end{eqnarray*}
The normalization constant $C$ is determined by setting $\zeta=x$ and
using the properties (\ref{eqn:inverse1}).
\begin{eqnarray*}
&&C=(x^2;x^4)_\infty g^{\rm R}~ \tr_{\Hr}\left(x^{2D}\right)
={(x^4;x^4,x^4)_\infty \over (x^6;x^4,x^4)_\infty}.
\end{eqnarray*}
Since the solution of a
difference equation is determined only up to pseudo-constants,
it is necessary to verify (e.g. by examining the analyticity)
that the above solution does give the quantity defined through the trace.
We have not proved this but checked it to some order in $x$.

\subsection{8 vertex model at the Ising point}
As is well known, the 8 vertex model at a special value of the parameters
can be regarded as the superposition of two non-interacting Ising models.
Let us consider a checkerboard lattice whose faces are shaded
alternatingly.
Place two kinds of Ising spins on the faces, one on the shaded faces
and the other on the unshaded ones.
In both cases we assume the $+$ boundary condition.
Place also 8 vertex `spin' variables  on the edges taking values $\pm $.
Consider a vertical (resp. horizontal)
 edge separating two faces that carry the spin variables
$\sigma_1$, $\sigma_2$.
We let the edge varibale $\vep$ have  the value $\sigma_1\sigma_2$ if
the left (resp.upper) face is shaded, and $-\sigma_1\sigma_2$ if otherwise.
In this way the Ising configurations are mapped to the 8 vertex configurations
with anti-ferroelectric boundary conditon.

The space of states for this doubled model is
the tensor product of the Fock spaces
$\H^0=\Hn\otimes \Hr$,
$\H^1=\Hr\otimes \Hn$.
This can be seen by combining two Ising configurations
to form an 8-vertex configuration:
the Ising configurations must belong to two different
sectors.

The following operators act on the doubled space $\H^0$:
$\phi^{\rm N}_r \otimes {\rm id}, \psi^{\rm NS}_0 \otimes \phi^{\rm R}_r$,
which are mutually anti-commuting.
For brevity, we refer to them by
$\phi^{\rm N}_r, \phi^{\rm R}_r$, respectively.
Similar conventions apply to $\H^1$.

For $i=0,1\in\Z/2\Z$, we define the VO's
$\Phi^{i+1}_{i\vep}:\H^i\rightarrow \H^{i+1}$
by
\begin{eqnarray}
&&\Phi_{0\vep}^{1}(\zeta)
=\sum_{\sigma}
\Phi_{\rN,\sigma}^{\rR V}(\zeta)
\otimes \Phi_{\rR}^{\rN,-\vep\sigma, V}(\zeta), \n
&&\Phi_{1\vep}^{0}(\zeta)
=\sum_{\sigma}
\Phi_{\rR}^{\rN,\sigma,V}(\zeta)
\otimes \Phi_{\rN,\vep\sigma}^{\rR, V}(\zeta). \label{eqn:8v}
\end{eqnarray}
Using the commutation relations for the Ising VO's we find that
(\ref{eqn:8v}) satisfies
\[
\Phi^i_{i+1\vep_2}(\zeta_2)\Phi^{i+1}_{i\vep_1}(\zeta_1)
=\sum_{\vep_1',\vep_2'}R(\zeta_1/\zeta_2)_{\vep_1\vep_2;\vep_1'\vep_2'}
\Phi^i_{i+1\vep'_1}(\zeta_1)\Phi^{i+1}_{i\vep'_2}(\zeta_2).
\]
The $R$ matrix is the one given in \cite{JMN} specialized to $q=x^4$,
except that our $\zeta^{-1}$ is the $\zeta=\zeta_{8v}$ there; explicitly,
in the notation of \cite{JMN},
\begin{eqnarray*}
{a(\zeta_{8v})\over \kappa(\zeta_{8v})}&=&w_-(\zeta)\bw_+(\zeta), \quad
{b(\zeta_{8v})\over \kappa(\zeta_{8v})}=w_+(\zeta)\bw_-(\zeta), \\
{c(\zeta_{8v})\over \kappa(\zeta_{8v})}&=&w_+(\zeta)\bw_+(\zeta), \quad
{d(\zeta_{8v})\over \kappa(\zeta_{8v})}=w_-(\zeta)\bw_-(\zeta).
\end{eqnarray*}

The matrix elements of VO's can also be obtained in a straightforward way.
We give below the generating functions of the one- and two-particle
matrix elements using
\[
\phi(z)=\phi^\rN(z)+\phi^\rR(z)=\sum_{r\in {1\over 2}\Z}\phi_r z^{-r}.
\]
Set $\zeta=1$, and write $\Phi_\vep=\Phi_{0\, -\vep}^{1}(1)$
or $\Phi_{1\, \vep}^{0}(1)$.
\begin{eqnarray}
&&
\br{\Phi_+\phi(z)\phi(w)}=
{(x^2z)_\infty \over (z)_\infty}
{(x^2w)_\infty \over (w)_\infty}
{(z-w)(1-x^{-1}\sqrt{zw}) \over
(\sqrt{z}-x \sqrt{w})(\sqrt{z}-x^{-1}\sqrt{w})},\\
&&\br{\Phi_-\phi(z)\phi(w)}=
ix^{-1/2}(\sqrt{z}-\sqrt{w})
{(x^2z)_\infty \over (z)_\infty}
{(x^2w)_\infty \over (w)_\infty}, \\
&&\br{\phi(z)\Phi_+\phi(w)}=
(1+{\sqrt{w}\over \sqrt{z}})
{(x^4z^{-1})_\infty \over (x^2z^{-1})_\infty}
{(x^2w)_\infty \over (w)_\infty}, \\
&&\br{\phi(z)\Phi_-\phi(w)}=
-ix^{-1/2}{1\over \sqrt{z}}
{(x^4z^{-1})_\infty \over (x^2z^{-1})_\infty}
{(x^2w)_\infty \over (w)_\infty}
{(z-w)(x-\sqrt{zw}) \over
(\sqrt{z}-x\sqrt{w})(\sqrt{z}-x^{-1}\sqrt{w})},\nonumber\\
&&\\
&&
\br{\phi(z)\phi(w)\Phi_+}=
-{1 \over \sqrt{zw}}
{(x^4z^{-1})_\infty \over (x^2z^{-1})_\infty}
{(x^4w^{-1})_\infty \over (x^2w^{-1})_\infty}
{(z-w)(x-\sqrt{zw}) \over
(\sqrt{z}-x\sqrt{w})(\sqrt{z}-x^{-1}\sqrt{w})},\nonumber\\
&&\\
&&
\br{\phi(z)\phi(w)\Phi_-}=
-ix^{1/2}{\sqrt{z}-\sqrt{w}\over \sqrt{zw}}
{(x^4z^{-1})_\infty \over (x^2z^{-1})_\infty}
{(x^4w^{-1})_\infty \over (x^2w^{-1})_\infty}.
\end{eqnarray}
Here we used $(z)_\infty=(z;x^4)_\infty$.


\def\Phih{\widehat{\Phi}}

\subsection{Ising model in the RSOS formulation}
In this subsection we examine the Ising model formulated as an ABF model
with $L=4$. Using the representation theoretical
construction
we shall calculate the two point functions for the VOs.

For our purpose it is useful to study first the symmetry under the Dynkin
diagram automorphism.
Let $\nu$ denote the algebra automorphism of $U'$ given by
$\nu(e_i)=e_{1-i}$, $\nu(f_i)=f_{1-i}$, $\nu(t_i)=t_{1-i}$. For
$\lambda=(k-l)\Lambda_0+l\Lambda_1$ set $\bar{\lambda}=
l\Lambda_0+(k-l)\Lambda_1$. Then there exists a unique isomorphism
(which we denote also by $\nu$) of vector spaces
$V(\lambda)\rightarrow V(\bar{\lambda})$ such that $\nu(\ket{\lambda})
=\ket{\bar{\lambda}}$ and $\nu(x\ket{u})=\nu(x)\nu(\ket{u})$ for all
$x\in U'$ and $\ket{u}\in V(\lambda)$. To make manifest the symmetry under
$\nu$ let us introduce the VO of vertex type in the principal picture:
\[
\Phih^{\mu\,V}_{\lambda}(\zeta)=
\zeta^{(\mu-\lambda,\rho)}\times
\bigl({\rm id}\otimes \zeta^{\bar{\rho}}\bigr)
\Phit^{\mu\,V}_{\lambda}(\zeta^2)
\]
where $\Phit^{\mu\,V}_{\lambda}(z)$ is the VO (\ref{eqn:typeI}).
In our case $\bar{\rho}=\Lambda_1-\Lambda_0$. The power of $\zeta$
is supplied in order to retain the normalization
$\Phih^{\lambda_{l\pm1}\,V}_{\lambda_l}(\zeta)\ket{\lambda_l}=
\ket{\lambda_{l\pm 1}}\otimes v_{\mp}+O(\zeta)$. We have
\[
\Phih^{\bar{\mu}\,V}_{\bar{\lambda}}(\zeta)
=\bigl(\nu\otimes \sigma^x\bigr)\circ
\Phih^{\mu\,V}_{\lambda}(\zeta)\circ\nu^{-1}.
\]
\def\Hom{{\rm Hom}}
Given
$f\in \Hom_{U'}\bigl(V(\lambda),V(\xi)\otimes V(\eta)\bigr)$
we write $\nu(f)=(\nu\otimes \nu)\circ f\circ\nu^{-1}
\in \Hom_{U'}\bigl(V(\bar{\lambda}),V(\bar{\xi})\otimes V(\bar{\eta})\bigr)$.
Define the VO of face type in the principal picture
$\Phih^{(\xi;\eta'\eta)}_{\lambda'\lambda}(\zeta)$ by
\begin{equation}
\left(\id_{V(\xi)}\otimes \Phih^{\eta' V}_{\eta}(\zeta)\right)\circ f
=\sum_{\lambda'}\left(\Phih^{(\xi;\eta'\eta)}_{\lambda'\lambda}
(\zeta)(f)\otimes\id\right)\circ
\Phih^{\lambda'V}_{\lambda}(\zeta).
\label{eqn:princi}
\end{equation}
Then we have the symmetry
\[
\Phih^{(\bar{\xi};\bar{\eta'}\bar{\eta})}_{\bar{\lambda'}\bar{\lambda}}(\zeta)(f)
=\nu\left(\Phih^{(\xi;\eta'\eta)}_{\lambda'\lambda}(\zeta)
(\nu^{-1}(f))\right).
\]

Now we focus attention to the Ising model. Taking $k=2$, we write
$\Phih^{(j;ii')}_{ll'}(\zeta)$ for
$\Phih^{(\Lambda_j;\Lambda_i\Lambda_{i'})}_{\lambda_l\lambda_{l'}}
(\zeta)$ (recall that $\lambda_l=(2-l)\Lambda_0+l\Lambda_1$).
Exhibiting the $j$ dependence explicitly we set
\[
\H^{(j;i)}_l=\{v\in V(\Lambda_j)\otimes V(\Lambda_i)\mid
e_sv=0,~ t_sv=q^{\br{h_s,\lambda_l}}v\quad (s=0,1)\},
\]
so that
\[
V(\Lambda_j)\otimes V(\Lambda_i)=
\bigoplus_{l\equiv i+j \,mod\, 2}\H^{(j;i)}_l\otimes V(\lambda_l).
\]
Clearly $\nu$ induces the isomorphism $\H^{(j;i)}_l\simeq
\H^{(1-j;1-i)}_{2-l}$.
We shall identify
\[
\H^{(0;0)}_0=\H^{\rN\,+},\quad
\H^{(0;0)}_2=\H^{\rN\,-},\quad
\H^{(0;1)}_1\simeq \H^{(1;0)}_1=\H^{\rR}
\]
and set
\begin{eqnarray*}
&&\vac_{\rN}=\ket{\Lambda_0}\otimes\ket{\Lambda_0}\in \H^{(0;0)}_0,\\
&&\vac_{\rR,i}=\ket{\Lambda_i}\otimes\ket{\Lambda_{1-i}}
\in \H^{(i;1-i)}_{1}.
\end{eqnarray*}
Defining the dual vectors in a similar manner, we wish to compute the
two point functions of $\Phih^{(j;ii')}_{ll'}(\zeta)$.
{}From their defining property (\ref{eqn:princi}) it follows that
\begin{eqnarray}
&&
\bra{\Lambda_0}\Phih^{\Lambda_0V}_{\Lambda_1}(\zeta_1)
\Phih^{\Lambda_1V}_{\Lambda_0}(\zeta_2)\ket{\Lambda_0}\nonumber\\
&&={}_{\rN}\dvac
\Phih^{(0;01)}_{01}(\zeta_1)
\Phih^{(0;10)}_{10}(\zeta_2)\vac_{\rN}
\times
\bra{\lambda_0}
\Phih^{\lambda_0V}_{\lambda_1}(\zeta_1)
\Phih^{\lambda_1V}_{\lambda_0}(\zeta_2)
\ket{\lambda_0}
\label{eqn:func1}\\
&&
\bra{\Lambda_i}\Phih^{\Lambda_iV}_{\Lambda_{1-i}}(\zeta_1)
\Phih^{\Lambda_{1-i}V}_{\Lambda_i}(\zeta_2)\ket{\Lambda_i}
=\sum_{l=0,2}
{}_{\rR,{1-i}}\dvac
\Phih^{(1-i;i1-i)}_{1l}(\zeta_1)
\Phih^{(1-i;1-ii)}_{l 1}(\zeta_2)
\vac_{\rR,{1-i}}\nonumber\\
&&\times
\bra{\lambda_1}
\Phih^{\lambda_1V}_{\lambda_l}(\zeta_1)
\Phih^{\lambda_l V}_{\lambda_1}(\zeta_2)
\ket{\lambda_1}.
\label{eqn:func2}
\end{eqnarray}

On the other hand,
the following can be obtained by solving the $q$KZ equation
for VOs of vertex type.
\begin{eqnarray*}
&&\bra{\Lambda_i}
\Phih^{\Lambda_iV}_{\Lambda_{1-i}}(\zeta_1)
\Phih^{\Lambda_{1-i}V}_{\Lambda_{i}}(\zeta_2)\ket{\Lambda_i}
=
{(x^6\zeta^2;x^4)_\infty\over(x^4\zeta^2;x^4)_\infty}
\left(v_{\pm}\otimes v_{\mp}+x\zeta v_{\mp}\otimes v_{\pm}\right)
\\
&&\bra{\lambda_l}
\Phih^{\lambda_l V}_{\lambda_{1}}(\zeta_1)
\Phih^{\lambda_1V}_{\lambda_l}(\zeta_2)
\ket{\lambda_l}
=
{(x^8\zeta^2;x^4,x^4)_\infty\over
(x^6\zeta^2;x^4,x^8)_\infty(x^{14}\zeta^2;x^4,x^8)_\infty}
\left(v_{\pm}\otimes v_{\mp}+x\zeta v_{\mp}\otimes v_{\pm}\right)
\\
&&
\bra{\lambda_1}
\Phih^{\lambda_1V}_{\lambda_l}(\zeta_1)
\Phih^{\lambda_l V}_{\lambda_{1}}(\zeta_2)
\ket{\lambda_1}
=
{(x^8\zeta^2;x^4,x^4)_\infty
\over(x^6\zeta^2;x^4,x^8)_\infty(x^{14}\zeta^2;x^4,x^8)_\infty}
\times\nonumber\\
&&
\left(
\phi\left({{1/4~-1/4}\atop {1/2}};x^8,x^{10}\zeta^2\right)
v_{\mp}\otimes v_{\pm}
+{x^3\zeta \over 1+x^2}
\phi\left({{1/4~3/4}\atop {3/2}};x^8,x^{10}\zeta^2\right)
v_{\pm}\otimes v_{\mp}\right).
\end{eqnarray*}
Here $\zeta=\zeta_2/\zeta_1$, and the sign $\pm$ is chosen according as
$i=0,1$ or $l=0,2$, respectively.
Using the symmetry
\[
{}_{\rR,{1}}\dvac
\Phih^{(1;01)}_{1l}(\zeta_1)
\Phih^{(1;10)}_{l 1}(\zeta_2)
\vac_{\rR,{1}}
=
{}_{\rR,{0}}\dvac
\Phih^{(0;10)}_{12-l}(\zeta_1)
\Phih^{(0;01)}_{2-l 1}(\zeta_2)
\vac_{\rR,{0}},
\]
we can solve (\ref{eqn:func1}), (\ref{eqn:func2})
for the two point functions of the $\Phih^{(j;ii')}_{ll'}(\zeta)$.
With the identificaiton
\begin{eqnarray*}
&&\Phih^{(0;01)}_{01}(\zeta)=\Phi^+(\zeta^{-1}),\quad
\Phih^{(0;01)}_{21}(\zeta)=\Phi^-(\zeta^{-1}),\\
&&\Phih^{(0;10)}_{10}(\zeta)=\Phi_+(\zeta^{-1}),\quad
\Phih^{(0;10)}_{12}(\zeta)=\Phi_-(\zeta^{-1}),
\end{eqnarray*}
we then recover the formulas
given in (\ref{eqn:vacexp1}), (\ref{eqn:vacexp2}).

\def\eq{\begin{eqnarray}}
\def\endeq{\end{eqnarray}}
\def\n{\nonumber\\}
\def\nn{\nonumber}
\def\vphi{\varphi}
\def\vac{|{\rm vac}\rangle}
\def\lvac{\langle {\rm vac}|}
\def\i{\iota}
\def\b{\bullet}
\def\hphi{{\hat \phi}}
\def\hvphi{{\hat \varphi}}
\def\End{{\rm End}}
\def\Hom{{\rm Hom}}
\def\H{{\cal H}}
\def\F{{\cal F}}

\subsection{Creation and annihilation operators}
To conclude this section we shall briefly discuss the
creation and annihilation operators for the row-to-row Hamiltonian.
We shall introduce two kinds of such operators,
and discuss the relation between them.

So far we have chosen to work with the $+$ boundary condition for the CTM.
To take the excited states into account, it is necessary to consider both of
the two boundary conditions $\sigma_j=\pm$ $(j\gg 1)$.
We denote the space for the CTM with those boundary conditions by $\H_\pm$
respectively, where $\H$ refers to either $\H^R$ or $\H^{NS}$.
The two spaces are identified by $\i=\prod_{\ell>0} \sigma_\ell^x$
(in the naive picture), which is the disorder operator at the origin.
The action of the VO's on $\H_-$ is defined by
$\Phi^\sigma(\zeta)|_{\H_{-}}=\i\circ\Phi^{-\sigma}(\zeta)\circ\i$
and  $\Phi_\sigma(\zeta)|_{\H_{-}}=\i\circ\Phi_{-\sigma}(\zeta)\circ\i$.

A priori there are $2\times2$ possible choices of the boundary conditions
for the full space on which the row-to-row Hamiltonian acts.
For simplicity we shall fix the boundary condition for the left half
to be $+$, and set $\F_{+,\pm}=\Hom(\H_\pm,\H_+)\simeq\H_+\otimes\H_\pm^*$.
The dual space is
$\F_{+,\pm}^{*}=\Hom(\H_+,\H_\pm)\simeq\H_+^{*}\otimes\H_\pm$
where the coupling
between $\F_{+,\pm}$ and $\F_{+,\pm}^{*}$
is chosen to be
$\langle f|g\rangle={\rm tr}_{\H_\pm}(fg)$ for
$f\in \F_{+,\pm}^*$ and $g\in \F_{+,\pm}$.
In this picture the ground state vector and its dual are $\vac=\lvac=x^{D}$.

The first way to introduce the creation and annihilation operators is
via the explicit diagonalization of the Hamiltonian using fermions.
Define
\[
\vphi(z)=\sum_m{1\over x^{2m}+x^{-2m}}{\phi_m\over z^m}.
\]
We find that the creation and annihilation operators
$\hvphi(z), \hvphi^*(z):\F_{\pm,+}\rightarrow \F_{\mp,+}$
are given by
\eq
&&\hvphi(z)(v)=\vphi(xz)\circ v\circ\i-v\circ\vphi(z/x)\circ\i,\n
&&\hvphi^*(z)(v)=\vphi(z/x)\circ v\circ\i+ v\circ \vphi(xz)\circ\i,
\endeq
where $v \in \F_{\pm,+}$.
They satisfy
$$
\hvphi(z)(\vac)=0,\quad
[\hvphi(z_1),\hvphi^*(z_2)]_+=\delta(z_1/z_2).
$$

The second way is to utilize
the transfer matrices $\tau^{R}_{NS}(\zeta)$, $\tau^{NS}_{R}(\zeta)$
in the representation theoretical picture,
as is done for the XXZ Hamiltonian and other models \cite{DFJMN},\cite{IIJMNT},
\cite{JMOh}.
Let us set
%
\begin{eqnarray*}
&&\tau^{R}_{NS}(\zeta)(v)=
(g^R)^{-1}\sum_\sigma \Phi(\zeta)_\sigma\circ v\circ\Phi(\zeta)^\sigma,
\\
&&\tau^{NS}_{R}(\zeta)(v)=
(g^{NS})^{-1}\sum_\sigma \Phi(\zeta)^\sigma\circ
v\circ\Phi(\zeta)_\sigma.
\end{eqnarray*}
If we define
$$
\hphi(z)(v)=\phi(z/x)\circ v\circ\i,
\quad \hphi^*(z)(v)=\phi(xz)\circ v\circ\i
$$
then we have the following commutaion relations:
%
\begin{eqnarray*}
&& \tau^{R}_{NS}(\zeta)\hphi^{NS}(z)=
f(z\zeta^2/x^2)\hphi^R(z)\tau^{R}_{NS}(\zeta),\\
&&\tau^{NS}_{R}(\zeta)\hphi^{R}(z)=
f(z\zeta^2/x^2)\hphi^{NS}(z)\tau^{NS}_{R}(\zeta).
\end{eqnarray*}
Here $f(z)$ signifies the function defined in Appendix A, (\ref{eqn:f(z)}).

The two types operators defined above are related, for example, as follows.
\eq
&&\hvphi^*(z)(\vac)=\hphi^*(z)(\vac)\n
&&\hvphi^*(z_1)(\hvphi^*(z_2)(\vac))=\hphi^*(z_1)(\hphi^*(z_2)(\vac))
-\delta(x^2z_1/z_2)\vac.\nn
\endeq

\bigskip
\noindent
{\sl Acknowledgement.}\quad
The authors would like to thank Fedor Smirnov
 for inspiring discussions.
One of the authors (M.J.) is grateful to Department of Mathematics, ANU
and Brian Davies for kind invitation and hospitality.
This work is partially supported by the Grant-in-Aid for Scientific Research
on Priority Areas, the Ministry of Education, Science and Culture, Japan,
and the Australian Research Council (ARC).

\appendix
\section{Matrix elements of the Ising VO}
In this appendix, we shall outline
the derivation of the matrix elements of VO's.

Let us consider $\Phi^{\rN,{\rm V}}_{\rR}(1)$.
Define $f(z)$ and $c_r$ by
\begin{eq}
&&f(z)=-\sqrt{k}{\rm sn}(v)
=-ix^{1/2}z^{-1/2}{\ip(z;x^4)\ip(x^4z^{-1};x^4)
\over\ip(x^2z;x^4)\ip(x^2z^{-1};x^4)}, \label{eqn:f(z)}
\end{eq}
where $v$ is related to $z$ by $z=exp(\pi iv/I)$.
Using the Fourier expansion
 $f(z)=\sum_{r\in{1\over2}+\Z}c_rz^{-r}$ on $|z|=1$
and the formulas (\ref{eqn:aNS}), (\ref{eqn:aR}) it can be shown that
\[
\sum_{j=1}^\infty a^\rN_{rj}a^\rR_{-s\,j+1}2\eta_s^{-1}=c_{r-s}.
\]
With the aid of this we rewrite the intertwining property (\ref{eqn:RNV})
as follows.
\[
\phi^\rN_r \Phi^{\rN,{\rm V}}_{\rR}(1)
=\sum_{s\in\Z} c_{r-s} \Phi^{\rN,{\rm V}}_{\rR}(1) \phi^\rR_s.
\]
In terms of the generating functions
\begin{eq}
\phi^\rN(z)&=&\sum_{r\in \Zh}\phi^\rN_r z^{-r}, \label{eqn:phiNS}\\
\phi^\rR(z)&=&\sum_{s\in \Z}\phi^\rR_s z^{-s}. \label{eqn:phiR}
\end{eq}
we have
$\phi^\rN(z) \Phi^{\rN,{\rm V}}_{\rR}(1)
=f(z) \Phi^{\rN,{\rm V}}_{\rR}(1) \phi^\rR(z)$,
or equivalently
\begin{equation}
\pt^\rN(z) \Phi^{\rN,{\rm V}}_{\rR}(1)
=\Phi^{\rN,{\rm V}}_{\rR}(1) \pt^\rR(z).
\label{eqn:intt}
\end{equation}
Here we have introduced auxiliary operators
\begin{eq}
\pt^\rN(z)&=&f_-(z)\phi^\rN(z)=\sum_{n\in \Z}\pt^\rN_n z^{-n}, \n
\pt^\rR(z)&=&f_+(z)\phi^\rR(z)=\sum_{n\in \Z}\pt^\rR_n z^{-n}
\end{eq}
according to the factorization $f(z)=f_+(z)/f_-(z)$, with
\[
f_+(z)={(z;x^4)_\infty \over (x^2z;x^4)_\infty},\quad
f_-(z)=ix^{-1/2}z^{1/2}{(x^2z^{-1};x^4)_\infty \over (x^4z^{-1};x^4)_\infty}.
\]
Notice that we have
\begin{eq}
&&\dvacn\pt^\rN_n=0\ (n<0),\n
&&\quad\pt^\rR_n\vacr=0\ (n>0),\quad
\quad\pt^\rR_0\vacr=\vacr.
\end{eq}
In view of (\ref{eqn:intt})
no confusion may arise if we abbreviate expressions like
\[
\dvacn \pt^\rN(z) \Phi^{\rN,{\rm V}}_{\rR}(1)\vacr
=\dvacn \Phi^{\rN,{\rm V}}_{\rR}(1) \pt^\rR(z)\vacr
\]
to $\dvacn \pt(z)\vacr$ and so on.
{}From the above properties, it follows that
\[
\dvacn \pt(z)\pt(w) \vacr=
{(z-w)\bigl(z+w-(1+x^{-2})zw\bigr)\over (z-x^2w)(z-x^{-2}w)}.
\]
Note that this formula contains
$\dvacn \pt(z)\vacr=1$ as a special case $w=0$.
The general matrix elements are given by Wick's theorem.

The other case $\Phi^{\rR,{\rm V}}_\rN(1)$ can be treated quite similarly.
The intertwining property becomes
\[
\phi^\rR(z)\Phi^{\rR,{\rm V}}_\rN(1)
=f(z)\Phi^{\rR,{\rm V}}_{\rN}(1)\phi^\rN(z)
\]
where $f(z)$ is the same as (\ref{eqn:f(z)}).
Defininig auxiliary operators
\[
\pt^{\rR'}(z)=f'_-(z)\phi^\rR(z),\quad
\pt^{\rN'}(z)=f'_+(z)\phi^\rN(z)
\]
with
\[
f'_+(z)=-ix^{1/2}z^{-1/2}{(z;x^4)_\infty \over (x^2z;x^4)_\infty},\quad
f'_-(z)={(x^2z^{-1};x^4)_\infty \over (x^4z^{-1};x^4)_\infty}
\]
we have
\begin{eq}
&&\pt^{\rR'}(z) \Phi^{\rR,{\rm V}}_{\rN}(1)
=\Phi^{\rR,{\rm V}}_{\rN}(1) \pt^{\rN'}(z),\n
&&
\dvacr\pt^{\rR'}_n=0\ (n<0),\quad
\dvacr\pt^{\rR'}_0=\dvacr,\qquad
\quad\pt^{\rN'}_n\vacn=0\ (n>0).\nonumber
\end{eq}
We find
\begin{eqnarray*}
\dvacr\pt'(z)\pt'(w)\vacn=
{(z-w)(z+w-1-x^2) \over (z-x^2w)(z-x^{-2}w)}.
\end{eqnarray*}

The matrix elements relative to the creation/annihilation operators in each
sector can be derived from the above generating functions.
Let $X_{mn}$ and $\gamma_m$ be as in (\ref{eqn:xgamma}).
Using the formula
\[
{(x^2z;x^4)_\infty \over(z;x^4)_\infty}=
\sum_{n=0}^\infty \gamma_n z^n
\]
we obtain
\def\issp{\Phi^{\rN,{\rm V}}_\rR(1)}
\begin{eqnarray*}
&&\dvacn\issp\pr_{-m}\pr_{-n}\vacr=X_{n,m}\gamma_m\gamma_n
\quad\quad (m,n\ge 0, m+n>0)\\
&&\dvacn\pn_{m+1/2}\issp\pr_{-n}\vacr=X_{-m-1/2,n}
ix^{2m+1/2}\gamma_m\gamma_n\ (m,n\ge0)\\
&&\dvacn\pn_{m+1/2}\pn_{n+1/2}\issp\vacr=
X_{m+1/2,n+1/2}x^{2m+2n+1}\gamma_m\gamma_n. \ (m,n\ge0)
\end{eqnarray*}
\def\ittp{\Phi^{\rR,{\rm V}}_{\rN}(1)}
Likewise we have
\begin{eqnarray*}
&&\dvacr \pr_{m}\pr_{n}\ittp\vacn=X_{m,n}x^{2m+2n}\gamma_m\gamma_n
\quad\quad (m,n\ge 0, m+n>0)\\
&&\dvacr\pr_{m}\ittp\pn_{-n-1/2}\vacn=
X_{m,-n-1/2}ix^{2m-1/2}\gamma_m\gamma_n\ (m,n\ge0)\\
&&\dvacr\ittp\pn_{-m-1/2}\pn_{-n-1/2}\vacn=
 X_{m+1/2,n+1/2} (-x^{-1}) \gamma_m\gamma_n.\ (m,n\ge0)\\
\end{eqnarray*}

Finally, by making use of the identity ($N$ even)
\[
{\rm Pfaffian}\left( {x_i-x_j \over 1-x_ix_j} \right)_{1\le i,j\le N}
=\prod_{1\le i<j\le N}{x_i-x_j \over 1-x_ix_j}
\]
we arrive at the general expressions (\ref{eqn:matele1}), (\ref{eqn:matele2}).

\section{Spin correlation functions}
The spin correlation functions for the Ising model are given as traces
of VO's.
They can be calculated using the intertwining property
and the unitarity of the VO's alone, without referring to the formulas for
the matrix elements.
In this appendix, we illustrate
this by the simplest case of the nearest diagonal
order-order and disorder-disorder correlations.

Let us introduce some notations. Setting $z=exp(\pi i v/I)$ we define
\begin{eqnarray*}
&&f^\rN_0(z)=\sqrt{2I k\over \pi}{\rm cn}(v)
=\sqrt{2\pi \over kI}\sum_{r\in \Zh}\eta^{-1}_r z^r,\n
&&f^\rR_0(z)=\sqrt{ 2I\over \pi}{\rm dn}(v)
=\sqrt{2\pi \over I}\sum_{s\in \Z}\eta^{-1}_s z^s. \n
\end{eqnarray*}
They are related to $f(z)=-\sqrt{k}{\rm sn}(v)$ given in (\ref{eqn:f(z)}) by
\[
f_0^\rN(x^2z)f(z)=if_0^\rR(z), \quad
f_0^\rR(x^{-2}z)f(z)=-if_0^\rN(z).
\]
Define further
\begin{eqnarray*}
\psi_1^\rN(\zeta)&=&
\zeta^{-D^\rN}\psi_1^\rN\zeta^{D^\rN}
=\oint{dz\over 2\pi i z} f_0^\rN(z)\phi^\rN(z/\zeta^2),\\
\psi_1^\rR(\zeta)&=&
\zeta^{-D^\rR}\psi_1^\rR\zeta^{D^\rR}
=\oint{dz\over 2\pi i z} f_0^\rR(z)\phi^\rR(z/\zeta^2).
\end{eqnarray*}
In terms of them and the generating functions
(\ref{eqn:phiNS}), (\ref{eqn:phiR}),
the intertwining properties of the VO's are summarized as follows:
\begin{eqnarray*}
\phi^\rN(z)\Phi^\sigma(\zeta)&=&
f(z\zeta^2)\Phi^{-\sigma}(\zeta)\phi^\rR(z), \n
\phi^\rR(z)\Phi_\sigma(\zeta)&=&
f(z\zeta^2)\Phi_{-\sigma}(\zeta)\phi^\rN(z), \n
\sigma \Phi^\sigma(\zeta)&=&\Phi^\sigma(\zeta)\psi_1^\rR(\zeta), \n
\sigma \Phi_\sigma(\zeta)&=&-i\Phi_{-\sigma}(\zeta)\psi_1^\rN(\zeta).
\end{eqnarray*}
Rewriting the properties for the horizontal VO's we have further
\begin{eqnarray*}
\Phi^\sigma(x\zeta)\phi^\rR(z)&=&
f(z\zeta^2)\phi^\rN(z)\Phi^{-\sigma}(x\zeta), \n
\Phi_\sigma(x\zeta)\phi^\rN(z)&=&
f(z\zeta^2)\phi^\rR(z)\Phi_{-\sigma}(x\zeta), \n
\sigma \Phi^\sigma(x\zeta)&=&i\psi_1^\rN(\zeta)\Phi^{-\sigma}(\zeta), \n
\sigma \Phi_\sigma(x\zeta)&=&\psi_1^\rR(\zeta)\Phi_{\sigma}(\zeta).
\end{eqnarray*}

The order-order and disorder-disorder correlators for the nearest diagonal
neighbors are given respectively by
\begin{eqnarray}
&&{\sum_{\sigma\sigma'}\sigma\sigma' \tr_{\H^\rN}
\left(x^{2D^\rN}\Phi^{\sigma'}(x\zeta_2)\Phi_\sigma(x\zeta_1)
\Phi^\sigma(\zeta_1)\Phi_{\sigma'}(\zeta_2)\right) \over
g^\rN g^\rR\tr_{\H^\rN}\bigl(x^{2D^\rN}\bigr)},\label{eqn:order}\\
&&{\sum_\sigma\tr_{\H^\rR}
\left(x^{2D^\rR}\Phi_{\sigma}(x\zeta_2)\Phi^\sigma(x\zeta_1)
\Phi_{-\sigma}(\zeta_1)\Phi^{-\sigma}(\zeta_2)\right) \over
g^\rN g^\rR\tr_{\H^\rR}\bigl(x^{2D^\rR}\bigr)}.
\label{eqn:disorder}
\end{eqnarray}
Here the $\zeta_i$ are the spectral parameters attached to the intermediate
lines as shown below.


Consider (\ref{eqn:order}). Using the intertwining properties and the
unitarity of the VO's we have
\[
\sum_\sigma \sigma \Phi_{\sigma}(x\zeta_1)\Phi^\sigma(\zeta_1)
=
\sum_\sigma \Phi_{\sigma}(x\zeta_1)\Phi^\sigma(\zeta_1)
\psi^\rR(\zeta_1)_1
=g^\rR \psi^\rR(\zeta_1)_1.
\]
Hence, setting $\zeta=\zeta_1/\zeta_2$, we find
\begin{eqnarray*}
&&\sum_{\sigma\sigma'}\sigma\sigma'
\Phi^{\sigma'}(x\zeta_2)\Phi_\sigma(x\zeta_1)
\Phi^\sigma(\zeta_1)\Phi_{\sigma'}(\zeta_2)\\
&&=g^\rR \sum_{\sigma'}\sigma' \Phi^{\sigma'}(x\zeta_2)\psi^\rR(\zeta_1)_1
\Phi_{\sigma'}(\zeta_2) \n
&&=g^\rR
\oint {dz\over 2\pi i z}f^\rR_0(z)
\sum_{\sigma'}\sigma' \Phi^{\sigma'}(x\zeta_2)\phi^\rR(z/\zeta_1^2)
\Phi_{\sigma'}(\zeta_2) \n
&&=g^\rR
\oint {dz\over 2\pi i z}f^\rR_0(z)f(z/\zeta^2)\phi^\rN(z/\zeta_1^2)
\sum_{\sigma'}\sigma' \Phi^{-\sigma'}(x\zeta_2)
\Phi_{\sigma'}(\zeta_2) \n
&&=-i g^\rR g^\rN
\oint {dz\over 2\pi i z}f^\rR_0(z)f(z/\zeta^2)\phi^\rN(z/\zeta_1^2)
\psi^\rN_1(\zeta_2).
\end{eqnarray*}
To take the trace we invoke the following simple lemma:
\begin{eqnarray*}
&&{\tr_{\Hr}\left(x^{2D^\rR}\phi^\rR(z_1)\phi^\rR(z_2)\right)
\over \tr_{\Hr}\bigl(x^{2D^\rR}\bigr)}=
\delta^\rR(x^2z_1/z_2),
\quad \delta^\rR(z)=\sum_{r\in\Z}z^r,\n
&&{\tr_{\Hn}\left(x^{2D^\rN}\phi^\rN(z_1)\phi^\rN(z_2)\right)
\over \tr_{\Hn}\bigl(x^{2D^\rN}\bigr)}=
\delta^\rN(x^2z_1/z_2),
\quad \delta^\rN(z)=\sum_{r\in{1\over2}+\Z}z^r.\\
\end{eqnarray*}
Here
the $\delta^\rN(z)$ contains half odd integer powers in $z$, but in the course
of the computation only integer powers appear.
With the aid of these formulas, (\ref{eqn:order}) becomes
\[
\oint{dz\over 2\pi i z}f_0^\rR(z) f_0^\rR(z/\zeta^2)
=
{2\pi \over I}\sum_{s\in \Z}\eta_s^{-2}\zeta^{2s}.
\]

The case of (\ref{eqn:disorder}) can be treated similarly.
We find that it is given by
\[
\oint{dz\over 2\pi i z}f_0^\rN(z) f_0^\rN(z/\zeta^2)
=
{2\pi \over kI}\sum_{r\in \Z+{1\over2}}\eta_r^{-2}\zeta^{2r}.
\]

\def\IJMPA{Int. J. Mod. Phys. A}
\def\CMP{Commun. Math. Phys.}
\def\LMP{Lett. Math. Phys.}
\def\NPB{Nucl. Phys. B}
\def\JSP{J. Stat. Phys.}
\def\JPC{J. Phys. C}
\def\PL{Phys. Lett.}
\def\RIMS{RIMS preprint}
\def\PA{Physica A}
\def\ICM{Proc. of the International Congress of Mathematicians, Kyoto}

\end{document}